\documentclass[prb,twocolumn,showpacs,floatfix,amsmath]{revtex4}

\usepackage{amssymb}
\usepackage{graphicx}
\usepackage{color}
\usepackage[normalem]{ulem}

\newcommand{\mrm}[1]{\mathrm{#1}}                     
\newcommand{\mcl}[1]{\mathcal{#1}}                    
\newcommand{\msf}[1]{\mathsf{#1}}                     

\newcommand{\ie}{\textit{i.e.}}
\newcommand{\eg}{\textit{e.g.}}
\newcommand{\cf}{\textit{cf.}}
\newcommand{\etal}{\textit{et al.}}
\newcommand{\apriori}{\textit{a priori}}
\newcommand{\viceversa}{\textit{vice versa}}
\newcommand{\abinitio}{\textit{ab initio}\ }

\newcommand{\transp}{\mrm{T}}                         
\newcommand{\vcrm}[1]{\boldsymbol{\mathrm{#1}}}       
\newcommand{\matx}[1]{\mcl{#1}}                       
\newcommand{\unmx}[1]{\matx{\mathds{1}}}              
\newcommand{\trace}[1]{\msf{tr}\!\left[#1\right]}     
\newcommand{\sign}[1]{\msf{sign}\!\left(#1\right)}    
\newcommand{\abs}[1]{\lvert#1\rvert}                  

\newcommand{\imagi}{\mathrm{i}}                       

\newcommand{\tdd}[2]{\frac{\mrm{d} #1}{\mrm{d} #2}}   
\newcommand{\intover}[2]{\mrm{d}^{#1} #2}             

\makeatletter
\newcommand\sh@ftsym[1]{%
  \smash{\raise-.3ex\hbox{$\scriptscriptstyle#1$}}}
\newcommand\mP{\mathbin{\sh@ftsym({\mp}\sh@ftsym)}}
\makeatother

\makeatletter
\newcommand\pM{\mathbin{\sh@ftsym({\pm}\sh@ftsym)}}
\makeatother
\newcommand{\curl}{\mathop{\mathrm{curl}}\nolimits}

\newcommand{\A}{A}
\newcommand{\Dvec}{\vcrm{D}}
\newcommand{\Kmatx}{\mcl{K}}

\newcommand{\beq}{\begin{equation}}
\newcommand{\eeq}{\end{equation}}
\newcommand{\bea}{\begin{eqnarray}}
\newcommand{\eea}{\end{eqnarray}}
\newcommand{\komma}{\, \mathrm{,}}
\newcommand{\punkt}{\, \mathrm{.}}

\newcommand{\einheit}[2]{#1~#2}
\newcommand{\Angs}{\mbox{\r{A}}}
\newcommand{\fleur}{\textsc{fleur} }

\makeatother

\begin{document}

\title{Role of Dzyaloshinskii-Moriya interaction for magnetism in transition-metal chains at Pt step-edges} 
\author{B.~Schweflinghaus}
\author{B.~Zimmermann}
\author{M.~Heide}
\author{G.~Bihlmayer}
\author{S.~Bl\"ugel}
\affiliation{Peter Gr\"unberg Institut and Institute for Advanced Simulation, Forschungszentrum J\"ulich and JARA, 52425 J\"ulich, Germany}
\date{\today}
\begin{abstract}
 We explore the emergence of chiral magnetism in one-dimensional monatomic Mn, Fe, and Co chains deposited at the Pt(664) step-edge carrying out an \abinitio study based on density functional theory (DFT). 
 The results are analyzed employing several models: 
 (i)  a micromagnetic model, which takes into account the Dzyaloshinskii-Moriya interaction (DMI) besides the spin stiffness and the magnetic anisotropy energy, and 
 (ii) the Fert-Levy model of the DMI for diluted magnetic impurities in metals. 
 Due to the step-edge geometry, the direction of the Dzyaloshinskii vector ($\Dvec$-vector) is not predetermined by symmetry and points in an off-symmetry direction. 
 For the Mn chain we predict a long-period cycloidal spin-spiral ground state of unique rotational sense on top of an otherwise atomic-scale antiferromagnetic phase. 
 The spins rotate in a plane that is tilted relative to the Pt surface by $62^\circ$ towards the upper step of the surface. 
 The Fe and Co chains show a ferromagnetic ground state since the DMI is too weak to overcome their respective magnetic anisotropy barriers. 
 An analysis of domain walls within the latter two systems reveals a preference for a Bloch wall for the Fe chain and a N\'eel wall of unique rotational sense for the Co chain in a plane tilted by $29^\circ$ towards the lower step. 
 Although the atomic structure is the same for all three systems, not only the size but also the direction of their effective $\Dvec$-vectors differ from system to system. 
 The latter is in contradiction to the Fert-Levy model. 
 Due to the considered step-edge structure, this work provides also insight into the effect of roughness on DMI at surfaces and interfaces of magnets. 
 Beyond the discussion of the monatomic chains we provide general expressions relating \abinitio results to realistic model parameters that occur in a spin-lattice or in a micromagnetic model. 
 We prove that a planar homogeneous spiral of classical spins with a given wave vector rotating in a plane whose normal is parallel to the $\Dvec$-vector is an exact stationary state solution of a spin-lattice model for a periodic solid that includes Heisenberg exchange and DMI. 
 In the vicinity of a collinear magnetic state, assuming that the DMI is much smaller than the exchange interaction, the curvature and slope of the stationary energy curve of the spiral as function of the wave vector provide directly the values of the spin stiffness and the spiralization required in micromagnetic models. 
 The validity of the Fert-Levy model for the evaluation of micromagnetic DMI parameters and for the analysis of \abinitio calculations is explored for chains. 
 The results suggest that some care has to be taken when applying the model to infinite periodic one-dimensional systems.
\end{abstract}

\maketitle

\section{Introduction}

In a seminal work, Gambardella \etal\cite{GDM+2002,Gamb2003} showed for the first time the presence of a truly one-dimensional (1D) metallic magnet. 
They succeeded in growing high-density arrays of monatomic Co chains on vicinal Pt(997) surfaces,\cite{DCE+2000,GBB+2000,GBB+2000b} denoted as Co/Pt(997), and investigated the magnetic properties by X-ray magnetic circular dichroism (XMCD). 
They found that, below a blocking temperature of about $T_\mrm{B}=\einheit{15}{\mrm{K}}$, a long-range ordered collinear spin state is observed with magnetic moments aligned in the easy axis direction. 
The authors explained this ferromagnetic order with a large magnetic anisotropy energy (MAE) of $\Delta E=\einheit{(2.0\pm0.2)}{\mrm{meV~/~Co~atom}}$ that counteracts the magnetic fluctuations due to the finite temperature. 
The success in growing and measuring 1D magnetic monatomic chain structures as well as a detection of an unusual easy axis direction pointing perpendicular to the chain direction and tilted by an angle of $43^\circ$ towards the upper terrace triggered theoretical investigations based on density functional theory (DFT),\cite{KEDF2002,ShMP2004,ULS+2004,KoSF2006,BRBB2006,BBBR2006} that affirmed the presence of an unusual direction of the easy axis. 
The strong MAE could be traced back to the large spin-orbit coupling (SOC) contribution of the Pt substrate. 
Succeeding these pioneering experiments alternative 1D systems had been investigated, among those FePt alloys\cite{HonolkaPRL09} and submonolayer Fe stripes,\cite{HLK+2009,Carbone+2011} both on Pt(997), as well as Fe stripes\cite{Hammer:03} and Co zigzag chains,\cite{DupeNJP2015} both on an Ir(001) (5$\times$1) surface.

In this paper we address the question in how far these results and their interpretation remain unchanged in the light of the recently discovered interface induced Dzyaloshinskii-Moriya interaction (DMI).\cite{BHB+2007} 
The DMI\cite{Dzya1957,Mori1960} appears in magnetic systems that lack inversion symmetry and exhibit strong SOC. 
Only recently it was found to be an indispensable ingredient to understand non-collinear magnetic structures of unique rotational sense observed in thin films, for the first time demonstrated by Bode \etal\cite{BHB+2007} who measured and analyzed a Mn monolayer on a W(110) substrate. 
Up to now a number of similar systems are known in which the DMI leads to magnetic ground states that are described as cycloidal spin spirals\cite{SantosNewJPhys2008,FBV+2008,ZHBB2014} or to the formation of a two-dimensional generalization of spirals with one-dimensional propagation vectors, the topological magnetic skyrmions.\cite{HBM+2011,RommingScience2013} 
Also for biatomic Fe chains deposited on an Ir(001) (5$\times$1) surface such a DMI-induced non-collinear magnetic ground state has been predicted\cite{MoTH2009,MaTo2009} and experimentally verified shortly after.\cite{MMW+2012}

In the light of these analyses we turn to the Pt step-edge structure and investigate the leading magnetic interactions for different monatomic TM chains deposited along the step-edges. 
Due to the reduced symmetry occurring at step-edges, a complex interplay of DMI, MAE, and exchange interaction is found to determine the magnetic ground state or the rotation type within a domain wall.

The magnetic structures are explored in the context of a micromagnetic model that is introduced in Sec.~\ref{sec:chAnalyze}. 
There, we discuss consequences of the symmetry of the investigated structure on the magnetic anisotropy and DMI and derive two micromagnetic criteria that determine the appearance of homogeneous and inhomogeneous spin spirals as magnetic ground states. 
In Sec.~\ref{sec:chMethod} we give details on the unit cell and the performed DFT calculations. 
We proceed in Sec.~\ref{sec:chResults} with presenting the results of the performed calculations for the three investigated systems, monatomic chains of Mn, Fe, and Co at Pt(664) step-edges, and extract parameters for the previously discussed micromagnetic model. 
Based on these parameters we predict the magnetic ground state for each system and characterize possible domain wall structures. 
We conclude this paper with four appendices: 
In Appendix~\ref{app:micro-model} we relate the micromagnetic parameters to the parameters of a lattice-spin model. 
In Appendix~\ref{app:sss-spin-lattice-model} we show that the spin spiral as calculated from first principles is a stationary state of the lattice-periodic spin model containing Heisenberg interaction and DMI. 
In Appendix~\ref{app:sss-connection-parameter} we relate the micromagnetic parameters with the spin-spiral energetics as calculated from first principles.
In Appendix~\ref{app:sss-Fert-Levy-Model} we analyze the relation between the microscopic DM vectors as obtained from the Fert-Levy model and the micromagnetic DM vectors. 
DM vectors are evaluated and compared to the \abinitio results from the main text.

\section{Micromagnetic analysis of the step-edge structure}
\label{sec:chAnalyze}

\subsection{Symmetry considerations}
\label{sec:ConsSymmetry}

Many of the systems, in which the DMI is known to lead to a non-collinear magnetic ground state, consist of one or more layers of $3d$ transition-metal (TM) elements placed on top of a heavy element substrate\cite{BHB+2007,FBV+2008,ZHBB2014} and exhibit two mirror planes. 
This restricts the direction of easy, medium, and hard axis as well as the direction of the effective Dzyaloshinskii-vector\cite{HeBB2011} ($\Dvec$-vector) to high-symmetry directions. 
Thus, the $\Dvec$-vector always points along either easy, medium, or hard axis. 
In the step-edge structure discussed in this paper (see Fig.~\ref{fig:mod_stepedge}), however, only one mirror plane perpendicular to the chain direction remains. 
A consequence of this reduction of symmetry with respect to film structures is the previously mentioned easy axis direction for the Co chains, tilted by $43^\circ$ towards the upper terrace. 
Similarly, the rules of Moriya\cite{Mori1960} only allow to reduce the possible orientation of the $\Dvec$-vector to the plane perpendicular to the chain axis, which is why \abinitio calculations become necessary to determine not only the strength of the DMI but also the direction of the $\Dvec$-vector.

Due to this particular symmetry at hand, the search for the magnetic ground state takes place in a higher-dimensional space. 
Besides the strength of the $\Dvec$-vector and the differences among easy, medium, and hard axes, one has to include in the final analysis the relative angle between $\Dvec$ and the principal axes of the anisotropy tensor.

\subsection{The micromagnetic model}

\begin{figure}[tb]
 \centering
   \includegraphics[scale=1]{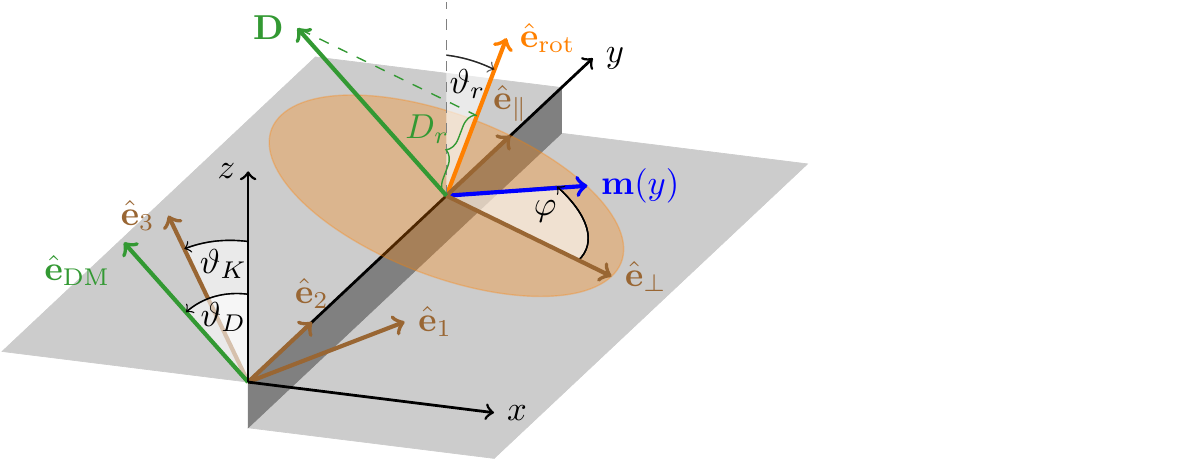}
 \caption{(color online)
          Step-edge structure, unit vectors and parameters 
          used in the text with respect to the Cartesian coordinate system ($x, y, z$): 
          The $\Dvec$-vector, being orthogonal to the $y$-axis, points along 
          $\hat{\vcrm{e}}_\mrm{DM}$ and encloses an angle $\vartheta_D$ with the $z$-axis. 
          The pairwise orthogonal principal axes of the anisotropy tensor $\Kmatx$, 
          $\hat{\vcrm{e}}_1$, $\hat{\vcrm{e}}_2$, and $\hat{\vcrm{e}}_3$, are associated 
          with $K_1$, $K_2$, and $K_3$, respectively, 
          where $\hat{\vcrm{e}}_3$ encloses an angle $\vartheta_K$ with the $z$-axis. 
          The rotation axis $\hat{\vcrm{e}}_\mrm{rot}$ is perpendicular to the $y$-axis and 
          encloses an angle $\vartheta_r$ with the $z$-axis. 
          $D_r$ denotes the projection of the $\Dvec$-vector onto $\hat{\vcrm{e}}_\mrm{rot}$. 
          $\vcrm{\hat{e}}_\parallel$ and $\vcrm{\hat{e}}_\perp$ are parallel and perpendicular to 
          the $y$-axis and are associated with $K_\parallel$ and $K_\perp$, respectively, 
          the anisotropy components within the rotation plane perpendicular to $\hat{\vcrm{e}}_\mrm{rot}$. 
          The magnetization density $\vcrm{m}(y)$ varies as function of distance along the step-edge ($y$-axis) 
          within the rotation plane (see semitransparent orange area) and encloses 
          the spin-spiral rotation angle $\varphi$ with $\vcrm{\hat{e}}_\perp$. 
          Note, that the angles $\vartheta_K$, $\vartheta_D$, and $\vartheta_r$ are positive (negative) 
          when pointing towards the lower (upper) terrace of the step-edge.}
 \label{fig:mod_stepedge}
\end{figure}

To systematically study the magnetic phases in a solid from first principles one usually employs a multiscale approach. 
DFT calculations are performed that allow to extract system-specific parameters, which characterize the behavior of the system in terms of a suitable model, \eg, a (generalized) Heisenberg or spin-lattice model\cite{LMBB2013} with spins placed on a discrete lattice. 
When the magnetic structure varies slowly across the crystal, meaning that the magnetic moments rotate on a length scale that is much larger than the interatomic distance, a micromagnetic model becomes favorable. 
Instead of a classical spin vector on each atomic site, such a model uses a continuous magnetization vector field $\vcrm{m}(\vcrm{r})$ (with $\vert\vcrm{m}\vert=1$) with effective parameters in which atom-specific contributions are implicitly contained. 
In case one deals with an antiferromagnetic spin-alignment, the classical spin vector is replaced by a staggered spin vector where the difference of up and down spins on neighboring atoms form a new order parameter, that is treated then as a continuous field. 
Regarding the atomic structure we deal with in this paper, a linear chain of magnetic atoms along the $y$ direction as depicted in Fig.~\ref{fig:mod_stepedge}, the magnetic energy for such spin textures can be expressed by the micromagnetic energy functional
\beq
    E[\vcrm{m}]
  = \int\intover{}{y} \left[
                             \frac{\A}{4\pi^2}\left(\dot{\vcrm{m}}\right)^2
                           + \frac{\Dvec}{2\pi}\cdot\left(\vcrm{m}\times\dot{\vcrm{m}}\right)
                           + \vcrm{m}^\transp\Kmatx\vcrm{m}
                      \right] \;,
  \label{eq:micmodEn}
\eeq
with $\vcrm{m}=\vcrm{m}(y)$ and $\dot{\vcrm{m}}=\frac{\mrm{d}}{\mrm{d}y}\vcrm{m}$.
The first term in Eq.~(\ref{eq:micmodEn}) contains the spin stiffness, $\A$, and favors collinear spins ($\dot{\vcrm{m}}\equiv0$). 
In contrast, the second term is linear in $\dot{\vcrm{m}}$ and thus shows a preference for a certain rotational sense of $\vcrm{m}$ with a strength and direction determined by the Dzyaloshinskii-vector, $\Dvec$. 
Finally, the magnetic anisotropy is accounted for by the last term, that features the anisotropy tensor, $\Kmatx$, whose principal axes point along hard, medium, and easy axes.\footnote{In general, one would need to include the non-local dipole-dipole interaction as well. Following the estimate in Ref.~\onlinecite{Bihl2005}, however, its contribution turns out to be negligible.} 
Note that in this work, without loss of generality, the energy-zero is given with respect to a magnetic configuration in which all spins are aligned along the easy axis. 
Furthermore, we point out that the local character of the integrand in Eq.~(\ref{eq:micmodEn}) is reasonable as long as the range of the magnetic interactions is shorter than the characteristic length scale of the magnetic structure that is described. 

In the following it is assumed to have knowledge of the model parameters $\A$, $\Dvec$, and $\Kmatx$. 
Considering the symmetry of the step-edge structure (\cf~Fig.~\ref{fig:mod_stepedge} and discussion in previous Sec.~\ref{sec:ConsSymmetry}) the latter two are of the form
\beq
  \Dvec  = \begin{pmatrix} D_x \\ 0 \\ D_z \end{pmatrix} \quad , \quad
  \Kmatx = \begin{pmatrix} K_{xx} & 0 & K_{xz} \\ 0 & K_{yy} & 0 \\ K_{xz} & 0 & K_{zz} \end{pmatrix} \punkt
  \label{eq:DvecKmatx_start}
\eeq
The direction of the Dzyaloshinskii vector, $\hat{\vcrm{e}}_\mrm{DM} = (\sin\vartheta_{\Dvec}, 0, \cos\vartheta_{\Dvec})^\transp$, is described with respect to the $z$-axis by the angle\footnote{In Eqs.~(\ref{eq:theta_D}) and (\ref{eq:theta_K}), we use $\mrm{atan2}\left(y,x\right)=\mrm{Arg}(x+\imagi y) \in \left(-180^\circ,180^\circ\right]$, the generalized form of the arcus-tangent function $\arctan\frac{y}{x} \in \left(-90^\circ,90^\circ\right)$, allowing to properly account for all sign combinations of $x$ and $y$.}
\beq
  \vartheta_{\Dvec} = \mrm{atan2}\left(D_x,D_z\right) \quad \in\ \left(-180^\circ,180^\circ\right] \punkt
  \label{eq:theta_D}
\eeq
The eigenvalues of the anisotropy ellipsoid $\Kmatx$, $K_1$, $K_2$, and $K_3$, are the magnetic anisotropies along the principal axes. 
The principal axis corresponding to $K_2$ is parallel to the $y$-axis. 
The axes $\hat{\vcrm{e}}_1=\matx{R}_y(\vartheta_{K}) \hat{\vcrm{e}}_x$ and $\hat{\vcrm{e}}_3=\matx{R}_y(\vartheta_{K}) \hat{\vcrm{e}}_z$ associated with $K_1$ and $K_3$ are obtained by a clockwise rotation, $\matx{R}_y(\vartheta_{K})$, of the magnetization $\vcrm{m}\rightarrow \matx{R}_y(\vartheta_{K})\vcrm{m}$ around the $y$-axis by an angle 
\beq
  \vartheta_{K} = \frac{1}{2} \mrm{atan2}\left(-2K_{xz},K_{xx}-K_{zz}\right) \quad \in\ \left(-90^\circ,90^\circ\right] \komma
  \label{eq:theta_K}
\eeq
which results to
\bea
  K_1 &=& K_{xx}\cos^2\vartheta_{K} - K_{xz}\sin2\vartheta_{K}+ K_{zz}\sin^2\vartheta_{K} \komma\\
  K_2 &=& K_{yy}                                                                          \komma\\
  K_3 &=& K_{xx}\sin^2\vartheta_{K} + K_{xz}\sin2\vartheta_{K}+ K_{zz}\cos^2\vartheta_{K} \punkt
\eea

\subsection{Homogeneous versus inhomogeneous flat spin spirals}
\label{sec:HomInhomSpSp}

It was first shown by Dzyaloshinskii\cite{Dzya1965} that the magnetization $\vcrm{m}(y)$ that minimizes the energy in functional~(\ref{eq:micmodEn}) may correspond to spins that are periodically modulated rather than collinearly aligned along the easy axis. 
According to the analysis of Heide \etal\cite{HeBB2011} such a non-collinear spin structure can be either a three-dimensional (3D) spin spiral or a flat spin spiral with a propagation vector $\vcrm{q}$ along the step-edge ($y$ direction) with magnetic moments rotating around the rotation axis of the spiral $\vcrm{\hat{e}}_\mrm{rot}=\left( \sin\vartheta_r , 0 , \cos\vartheta_r \right)^\transp$ with an angle $\vartheta_r$ that encloses $\vcrm{\hat{e}}_\mrm{rot}$ and the $z$-axis (the surface normal) and that is restricted to $-90^\circ<\vartheta_r\leq90^\circ$. 
In the following we restrict our analysis to flat spin spirals, \ie, the magnetization direction is always perpendicular to the rotation axis and $\vcrm{\hat{e}}_\mrm{rot}$ is independent of $y$. 
For one part, this allows an analytical treatment of the problem, and for the other, we will show in Sec.~\ref{sec:chResults} that for all investigated systems the regime of truly 3D spin spirals can be excluded. 
Thus, the magnetization direction along the chain is given by
\beq
  \vcrm{m}(y) = \matx{R}_y(\vartheta_{r})\begin{pmatrix} \cos\varphi(y) \\ \sin\varphi(y) \\ 0 \end{pmatrix}
\eeq
and depends on a 1D parameter, the spin-spiral rotation angle $\varphi(y)$. 
The matrix $\matx{R}_y$ describes a rotation around the $y$-axis. 
Inserting this into the energy functional from Eq.~(\ref{eq:micmodEn}), normalized to one period length one arrives at an expression for the average energy density,
\bea
      E_\lambda\left[\varphi,\vartheta_r\right]
  &=& \frac{1}{\lambda} \int_0^{2\pi} \!\!\!\! \mrm{d}\varphi
          \left[
                   \frac{\A}{4\pi^2}\dot{\varphi}
                 + \frac{K_\perp\cos^2\varphi + K_\parallel\sin^2\varphi}{\dot{\varphi}}
          \right] \nonumber\\
  & & + \frac{D_r}{\lambda} \komma
  \label{eq:micmodEn2}
\eea
with $\dot{\varphi}=\tdd{\varphi}{y}$. 
$D_r$ is the projection of the Dzyaloshinskii-vector onto the rotation axis and reads
\beq
  D_r = \Dvec\cdot\vcrm{\hat{e}}_\mrm{rot} = D_x\sin\vartheta_r + D_z\cos\vartheta_r \punkt
  \label{eq:Dr}
\eeq
$K_\perp$ and $K_\parallel$ denote the anisotropy components in the rotation plane of the magnetization perpendicular and parallel to the chain axis and are given by
\bea
  K_\perp     &=& K_{xx}\cos^2\vartheta_r - K_{xz}\sin2\vartheta_r + K_{zz}\sin^2\vartheta_r \komma \\
  K_\parallel &=& K_{yy} = K_2 \punkt
\eea
Note, that $D_r$ and $K_\perp$ depend explicitly on the angle of the rotation axis, $\vartheta_r$, such that the functional of the average energy density in Eq.~(\ref{eq:micmodEn2}) depends on $\vartheta_r$ as well. 
For later purposes we additionally define $K_\mrm{max}=\max\{K_\perp,K_\parallel\}$, $K_\mrm{min}=\min\{K_\perp,K_\parallel\}$, and the average $\overline{K}=\left(K_\perp+K_\parallel\right)/2$. 
Note, that $\lambda$ can become negative since the present formalism also accounts for the rotational sense of the spiral. 
We distinguish a right-rotating spiral for $\dot{\varphi}>0$ and $\lambda>0$ (energetically preferred when $D_r<0$, see Eq.~(\ref{eq:micmodEn2})) and a left-rotating spiral for $\dot{\varphi}<0$ and $\lambda<0$ (energetically preferred when $D_r>0$) following the convention that a left-rotating spiral rotates clockwise when projecting the magnetic moments onto the $xy$ plane and reading the spiral rotation along the positive $y$ direction (see Fig.~\ref{fig:mod_stepedge}).

For a homogeneous spin spiral $\varphi(y)$ changes linearly with distance within the chain and one finds $\varphi(y)=y\; \vcrm{q}\cdot\vcrm{\hat{e}}_y$, where $\vcrm{q}$ is the spin-spiral wave vector. 
Thus, $\dot\varphi=\vcrm{q}\cdot\vcrm{\hat{e}}_y=2\pi/\lambda=\mbox{const.}$, and Eq.~(\ref{eq:micmodEn2}) simplifies to
\beq
  E_\lambda^\mrm{hom} = \frac{\A}{\lambda^2} + \frac{D_r}{\lambda} + \overline{K} \komma
\eeq
\ie, the energy density shows a parabolic behavior with respect to the inverse of the spiral length. 
Only when the minimum of this expression,
\beq
    E_{\lambda_\mrm{min}}^\mrm{hom}
  = -\frac{D_r^2}{4\A}+\overline{K} \ , \ \mbox{with} \ \lambda_\mrm{min} = -2 \frac{\A}{D_r} \komma
  \label{eq:enehom}
\eeq
is below zero (corresponding to the energy of collinear spins aligned along the easy axis direction), a spiraling magnetic ground state can be established. 
This leads to the criterion for the appearance of a homogeneous spin spiral,
\beq
    \mrm{f}_\mrm{crit}^\mrm{hom}(\vartheta_r)
  = \frac{1}{4} \frac{D_r^2}{\A\overline{K}}
  \stackrel{!}{>} \ 1 \punkt
  \label{eq:Crit_hss}
\eeq
In the case of an inhomogeneous spin spiral ($\dot\varphi\neq\mbox{const.}$) the energy-density functional in Eq.~(\ref{eq:micmodEn2}) can be minimized by means of the Euler-Lagrange formalism\cite{Dzya1965} resulting in
\bea
  \label{eq:eneinh}
      E_\lambda^\mrm{inh}
  &=& -2\vert K_\perp-K_\parallel \vert
        \frac{\mrm{E}(\epsilon)}{\epsilon^2\mrm{K}(\epsilon)} - c
      + \frac{D_r}{\lambda^\mrm{inh}} \komma \\
      \lambda^\mrm{inh}
  &=& -\frac{2}{\pi}\sign{D_r}\sqrt{\A/\vert K_\perp-K_\parallel \vert}\; \epsilon\mrm{K}(\epsilon) \komma
  \label{eq:laminh}
\eea
with the Lagrange multiplier $c>-\min\{K_\perp,K_\parallel\}$. 
$\mrm{K}(\epsilon)$ and $\mrm{E}(\epsilon)$ are the complete elliptic functions of first and second kind,\footnote{The complete elliptic functions of first and second kind are defined by $\mrm{K}(\epsilon)=\int_0^{\pi/2}\mrm{d}\phi\;\left(1-\epsilon^2\sin^2(\phi)\right)^{-1/2}$ and $\mrm{E}(\epsilon)=\int_0^{\pi/2}\mrm{d}\phi\;\left(1-\epsilon^2\sin^2(\phi)\right)^{1/2}$, respectively.} respectively, with the ellipticity $\epsilon=\epsilon(c)=\sqrt{\vert K_\perp-K_\parallel \vert/(K_\mrm{max}+c)}$. 
It can be shown that the average energy density in Eq.~(\ref{eq:eneinh}) gets minimal when $E_\lambda^\mrm{inh}=-c$. 
Together with Eq.~(\ref{eq:laminh}) this leads to a conditional equation for $c=c(\epsilon)$,
\beq
  \vert D_r \vert = \frac{4}{\pi}\sqrt{\A\vert K_\perp-K_\parallel \vert} \; \frac{\mrm{E}(\epsilon)}{\epsilon} \punkt
  \label{eq:micmodDr}
\eeq
An inhomogeneous spiral appears for $-c<0$, leading to the criterion
\beq
    \mrm{f}_\mrm{crit}^\mrm{inh}(\vartheta_r)
  = \frac{1}{4} \frac{D_r^2}{\A\overline{K}}\cdot\alpha(K_\perp,K_\parallel) \stackrel{!}{>} 1 \komma
  \label{eq:Crit_iss}
\eeq
where
\beq
    \alpha(K_\perp,K_\parallel)
  = \frac{\overline{K}}{K_\mrm{max}}
    \left(\frac{2}{\pi}\mrm{E}\left(\sqrt{\frac{\vert K_\perp-K_\parallel\vert}{K_\mrm{max}}}\right)\right)^{-2}
\eeq
is a factor that depends on the ellipticity of the anisotropy energy within the plane of rotation of the magnetic moments spiral rotation axis: 
If the ellipticity $\epsilon$ within the rotation plane is zero ($K_\perp=K_\parallel$), then $\alpha=1$, which means that both criteria, Eqs.~(\ref{eq:Crit_hss}) and (\ref{eq:Crit_iss}), become identical. 
One can show that elsewise $\alpha>1$, meaning that the criterion for the appearance of an inhomogeneous spin spiral is always easier to be fulfilled than the criterion for the appearance of a homogeneous spiral, Eq.~(\ref{eq:Crit_hss}). 
For the case that the easy axis lies along the rotation axis ($K_\mrm{min}=0$) we have $\alpha=\pi^2/8$ and Eq.~(\ref{eq:Crit_iss}) simplifies to
\beq
  \frac{D_r^2}{\A K_\mrm{max}} \stackrel{!}{>} \frac{16}{\pi^2} \komma
\eeq
which has been already discussed in literature.\cite{Dzya1965,Izyu1984}

As a final remark we state that the anisotropy term in Eq.~(\ref{eq:micmodEn2}) can also be written as $\left(K_\perp-K_\parallel\right)\cos^2\varphi+K_\parallel$, which leads to the same expressions as derived above.

\subsection{Micromagnetic Parameters}
\label{sec:Micromagnetic Parameters}

The three micromagnetic parameters $\A$, $\Dvec$, and $\Kmatx$ are related to the site-dependent microscopic parameters of a spin-lattice model via
\bea
  \frac{\A}{4\pi^2}  &=& -\frac{1}{2\Delta} \sum_{j>0} R_{0j}^2 J_{0j}        \;, \nonumber\\
  \frac{\Dvec}{2\pi} &=&  \frac{1}{ \Delta} \sum_{j>0} R_{0j}   \vcrm{D}_{0j} \;,\quad\mbox{and}\quad
  \Kmatx              =   \frac{1}{ \Delta}                     \Kmatx_0      \;,
  \label{eq:ADK_connection}
\eea
where $\Delta$ defines the distance between two neighboring atoms within the chain, and $R_{0j}=j\,\Delta$ is the distance between atoms at sites $0$ and $j$.
$J_{0j}$, $\vcrm{D}_{0j}$, and $\Kmatx_0$ are the exchange interaction, the Dzyaloshinskii vector between a pair of atoms at sites $0$ and $j$, and the on-site anisotropy at the representative atom labeled $0$, respectively (see Appendix \ref{app:micro-model} and Ref.~\onlinecite{ZHBB2014} for details). 

The integrand of Eq.~(\ref{eq:micmodEn}) is an energy density. 
For the quasi one-dimensional magnets studied in this work, it has the unit \emph{energy per length}. 
Accordingly, the parameters $\A$, $\Dvec$, and $\Kmatx$ take the units \emph{energy times length}, \emph{energy}, and \emph{energy per length}, respectively. 
However, it is often convenient to use another normalization, and represent energy densities in units of \emph{energy per TM atom}. 
The conversion from the first normalization to the second one is done by multiplication with $\Delta$. 
In analogy, the units for the micromagnetic parameters $\A$, $\Dvec$, and $\Kmatx$ change to \emph{energy times area per TM atom}, \emph{energy times length per TM atom}, and \emph{energy per TM atom}, respectively. 
For the rest of this paper we use the same symbols for the two different normalizations, and the used normalization can be inferred from the unit.

Notice, it is customary that both communities, the micromagnetic and the spin-lattice model community, refer to $\vcrm{D}$ or $\vcrm{D}_{ij}$, respectively, as the Dzyaloshinskii-vector, although they are obviously different. 
We follow this tradition, but refer in addition to the $\vcrm{D}$-vector in the spin-lattice model as microscopic $\vcrm{D}$-vector, and in the micromagnetic model either as the micromagnetic or effective $\vcrm{D}$-vector or as the spiralization,\cite{FreimuthJPCM2014} whenever necessary.

The spin stiffness and spiralization can be obtained directly from first-principles calculations invoking the homogeneous spin-spiral state. 
In Appendix~\ref{app:sss-spin-lattice-model} we prove that for each wave vector $\vcrm{q}$ there are two flat homogeneous spin spirals of opposite handedness with a rotation axis parallel and antiparallel to the $\Dvec$-vector of that given mode. 
The lowest energy is found for a wave vector $\vcrm{Q}$ with a spin-chirality opposite to the $\Dvec$-vector of that mode. 
In Appendix~\ref{app:sss-connection-parameter} we show that if $\vcrm{Q}$ is in the vicinity of a high-symmetry point in the Brillouin zone, \eg, $\vcrm{Q}=\vcrm{0}$ in case of the ferromagnetic state, typically this implies that the DMI is small compared to the exchange interaction. 
For the step-edge structure $\vcrm{q}$ becomes one-dimensional and we obtain the spin stiffness from the curvature $A \propto 
\mrm{d}^2E(q)/\mrm{d}q^2$. 
The spiralization projected onto the direction of the DMI-vector is obtained from the slope $[\hat{\vcrm{e}}_\mrm{DM}\cdot\Dvec] \propto \mrm{d}E(q,\hat{\vcrm{e}}_\mrm{DM})/\mrm{d}q$ of the energy calculated for wave vectors $\vcrm{q}$ in the vicinity of the high-symmetry point. 
In the following Section we calculate $A$ and $\vcrm{D}$ from $E(q)$ for spin-spiral waves with $q$-vectors of different length from first principles in two separate steps: 
At first, $E(q)$ is calculated without spin-orbit interaction employing the generalized Bloch theorem, from which the spin stiffness is determined and for which the spiralization is zero by definition, and then the spiralization is determined by calculating the change of the total energy $\Delta E(q)$ adding the spin-orbit interaction in first order perturbation calculated from electronic states related to the spin-spiral solution.

\section{First-principles theory}
\label{sec:chMethod}

\subsection{Structural Model and Computational Details}

\begin{figure}[tb]
 \centering
   \includegraphics[scale=1]{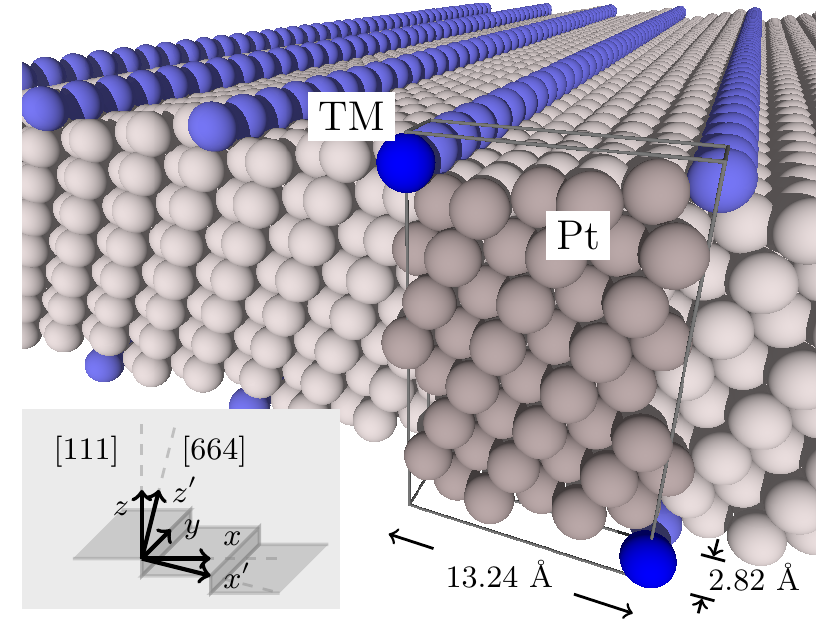}
 \caption{(color online)
          Sketch of the unit cell, a slab of a 
          (664) vicinal surface decorated with a monatomic chain 
          along the edge, and its repetition within the $x'y$ plane. 
          The dark blue spheres correspond to the transition metals (Co, Fe, Mn) 
          and the bright gray spheres represent the substrate atoms (Pt). 
          The chain-to-chain distance is \einheit{13.24}{\r{A}} and the nearest-neighbor distance 
          within the chain is \einheit{2.82}{\r{A}}. 
          The inset in the lower left illustrates the use of the coordinates within the text. 
          Note that the structure is periodic with respect to the $x'y$ plane. 
          Although in the actual calculation all quantities are referenced 
          with respect to $(x'yz')$, throughout this paper they are given 
          with respect to $(xyz)$, in accordance with Fig.~\ref{fig:mod_stepedge}.}
 \label{fig:unitcell}
\end{figure}

The \abinitio calculations based on DFT are carried out in film geometry of the full-potential linearized augmented plane-wave (FLAPW) method\cite{KrakauerPRB1979,WeinertRPB1981} as implemented in the \fleur code.\cite{fleur} 
The chains at stepped surfaces are modeled like in earlier studies,\cite{BRBB2006,BBBR2006} where the chosen unit cell, a (664) step-edge structure, turned out to be a suitable structural model. 
The setup of the unit cell is inversion symmetric and consists of a tilted 8-layer Pt slab with two monatomic TMs deposited on both sides of the slab onto the step-edge. 
Throughout this investigation no relaxation of the structure is considered. 
This is motivated by the finding that relaxations can lead to an unphysically strong quenching of the orbital moment and, thus, to less accurate results for the MAE.\cite{BRBB2006,BBBR2006,CoFB2008} 
In Fig.~\ref{fig:unitcell} we sketch the structural model and indicate the unit cell by the darker spheres in the foreground. 
The chain axis and, thus, the propagation direction of the investigated spiral structure is chosen as $y$-axis. 
To ensure a periodic repetition of the structure along the $x'$-direction, as required by a solid-state code, the steps of the surfaces with normal [111] direction are tilted by an angle of about $10^\circ$, so that the $z'$-direction of the unit cell is [664] and the $x'$-direction is $[11\overline{3}]$ (see inset in the lower left of Fig.~\ref{fig:unitcell}), resulting in a (6$\times$1) surface unit cell. 
The used lattice constant is $a_\mrm{Pt}=\einheit{3.99}{\Angs}$ as calculated by Baud \etal\cite{BRBB2006} 
Along the $x$ direction the unit cell has the length of the distance of two vicinal TM chains, $a_{x'}=\sqrt{11}\cdot a_\mrm{Pt}\approx\einheit{13.24}{\Angs}$. 
The width corresponds to the distance between two neighboring TM atoms within one chain, $a_y=a_\mrm{Pt} / \sqrt{2}\approx\einheit{2.82}{\Angs}$. 
For all types of atoms within the 2D unit cell the muffin-tin (MT) radius is chosen to be $R_\mrm{MT}=\einheit{1.16}{\Angs}$.
If not stated otherwise, all energies obtained from first-principles calculations refer to energies per computational unit cell. 
Depending then on the micromagnetic quantity under consideration, this energy can be related to energy per magnetic TM atom or per chain atom, respectively.

For the exchange and correlation functional we chose the local density approximation (LDA) as proposed by Moruzzi, Janak, and Williams.\cite{MoJW1978}
The computational cutoff values for the expansion of the Kohn-Sham potential are $\vcrm{G}_\mrm{max}=\einheit{12.0}{\mrm{a.u.}^{-1}}$ for the potential and $\vcrm{G}_\mrm{max}^\mrm{xc}=\einheit{9.5}{\mrm{a.u.}^{-1}}$ for the exchange-correlation potential. 
The Hamiltonian matrix elements for all atoms in the unit cell due to the non-spherical part of the potential are expanded up to $\ell_\mrm{max}^\mrm{n-sph}=6$. 
The spherical harmonics expansion of the LAPW basis includes functions up to $\ell_\mrm{max}=8$ within each MT sphere and all basis functions satisfying $\vert\vcrm{k}_\parallel+\vcrm{G}_\parallel\vert < K_\mrm{max}$ are included. 
If not stated otherwise, $K_\mrm{max}=\einheit{3.5}{\mrm{a.u.}^{-1}}$ is used. 
All self-consistent calculations have been carried out with 128 $\vcrm{k}$-points in the full 2D Brillouin zone, whereas for one-shot calculations employing the force theorem of Andersen\cite{MaAn1980} 512 $\vcrm{k}$-points have been used.

\subsection{Spin stiffness}

The parameters $\A$, $\Dvec$, and $\Kmatx$ are calculated as outlined in Refs.~\onlinecite{ZHBB2014} and \onlinecite{HeBiBl2009}. 
The spin stiffness, $\A$, is obtained by determining the total energy $E_{\text{SS}}(\vcrm{q})$ of the system as function of flat homogeneous spin spirals with wave vectors $\vcrm{q}$ of different lengths, all in the vicinity of the ferromagnetic or antiferromagnetic state and all along the chain direction. 
Since the lengths of the $\vcrm{q}$-vectors are small we applied the force theorem of Andersen\cite{MaAn1980} to obtain these energies as deviations from the collinear state whose densities are calculated self-consistently employing the scalar relativistic approximation and which served as the initial state from which the force theorem is applied. 
To avoid numerical errors the magnetization in the interstitial region was set to zero before applying the force theorem. 
A detailed description can be found in Ref.~\onlinecite{LMBB2013}. 
When calculating the spin stiffness we omitted the energy correction due to SOC, because it proved small in tests and thus we can restrict ourselves to the use of the generalized Bloch theorem.\cite{Sandratskii:91.1}

\subsection{Dzyaloshinskii-Moriya interaction}

The effective $\Dvec$-vector is determined treating SOC in first-order perturbation theory on top of flat homogeneous spin-spiral solutions used to determine $\A$. 
The DM energy is given by\cite{HeBiBl2009,ZHBB2014}
\beq
     E_\mrm{DM}(\vcrm{q},\hat{\vcrm{e}}_\mrm{rot})
   = \sum_{\vcrm{k}\nu}{ f(\epsilon^{0}_{\vcrm{k}\nu}, T) \,
     \delta{\epsilon}_{\vcrm{k}\nu}(\vcrm{q},\hat{\vcrm{e}}_\mrm{rot}) } \punkt
  \label{eq:EDM_smearingT}
\eeq
The occupation numbers are given by the Fermi function $f(\epsilon, T)$, which introduces a broadening of the occupation around the Fermi energy by the temperature $T$. 
They depend on the wave vector $\vcrm{q}$ through the unperturbed (\ie, without SOC) eigenvalue spectrum $\epsilon^{0}_{\vcrm{k}\nu}(\vcrm{q})$. 
The change of the eigenvalue spectrum
\beq
    \delta{\epsilon}_{\vcrm{k}\nu}(\vcrm{q},\hat{\vcrm{e}}_\mathrm{rot})
  = \langle   \matx{U}(\hat{\vcrm{e}}_\mathrm{rot}) \psi_{\vcrm{k}\nu}(\vcrm{q})
            | \mathcal{H}_\mathrm{so}
            | \matx{U}(\hat{\vcrm{e}}_\mathrm{rot}) \psi_{\vcrm{k}\nu}(\vcrm{q}) \rangle
  \label{eq:FstO_expctval}
\eeq
due to SOC described by the Hamiltonian $\mathcal{H}_\mathrm{so}$, depends additionally on the rotation axis $\hat{\vcrm{e}}_\mathrm{rot}$. 
The unitary transformation $\matx{U}(\hat{\vcrm{e}}_\mathrm{rot})$ directs the flat spin spiral of the unperturbed state rotating around the $z$-axis to the global spin-rotation axis and $\psi_{\vcrm{k}\nu}(\vcrm{q})$ denotes the spin-spiral eigenstates of the unperturbed Hamiltonian. 
The summation in Eq.~(\ref{eq:EDM_smearingT}) runs over all states characterized by the Bloch vector $\vcrm{k}$ and band index $\nu$. 
Due to the finite number of $\vcrm{k}$-points (512 $\vcrm{k}$-points in the whole 2D unit cell) the effect of the broadening temperature will be a subject of study in Sec.~\ref{sec:DzyaloshinskiiVector}, which allows an estimation for the qualitative reliability of our results.

We analyzed $E_\mrm{DM}$, the change of the DM energy for a set of $\vcrm{q}$-vectors that point along the chain direction (\ie, the $y$-axis) but vary in length, as well as two different rotation axes oriented along $x$- and $z$-direction ($\hat{\vcrm{e}}_\mrm{rot}=\hat{\vcrm{e}}_x$ and $\hat{\vcrm{e}}_\mrm{rot}=\hat{\vcrm{e}}_z$, see Fig.~\ref{fig:mod_stepedge} and Eq.~(\ref{eq:Dr})) to determine independently the two non-vanishing components of the $\Dvec$-vector (the third component vanishes due to symmetry, as already discussed in Sec.~\ref{sec:ConsSymmetry}). 
In the micromagnetic limit, \ie, in the limit of long-period spirals, Eq.~(\ref{eq:micmodEn2}) is applicable. 
Therefore, if the spin-orbit interaction is included, the DM energy is expected to change linearly with the length of the wave vector $\vcrm{q}$ in the vicinity of the collinear spin alignment. 
Consequently we evaluate the effective $\Dvec$-vector as the slope of the energy change with respect to $q$ in the limit $q\rightarrow 0$. 

As outlined in Ref.~\onlinecite{ZHBB2014}, the spin-orbit coupling operator $\mathcal{H}_\mathrm{so} $ can be safely approximated by an atom-by-atom superposition of SOC operators limited to the muffin-tin spheres of the atoms, \ie,
\beq
  \mathcal{H}_\mrm{so} = \sum_{\mu} \xi (r^{\mu}) \, \boldsymbol{\sigma} \cdot {\vcrm L}^{\mu} \komma
  \label{eq:SOCoperator}
\eeq
where $\xi$ is the spin-orbit strength related to the spherical muffin-tin potential $V(r^{\mu})$, $\xi \sim {r}^{-1} \, \mathrm{d}V/\mathrm{d}r$, ${\vcrm r}^{\mu} = {\vcrm r} - {\vcrm R}^{\mu}$, and $|{\vcrm r^{\mu}}| < R^{\mu}_{\mathrm{MT}}$. 
${\vcrm R}^{\mu}$ references the center and $R^\mu_\mathrm{MT}$ is the radius of the $\mu$th muffin-tin sphere, with $\mu$ running over all atoms in the unit cell. 
The atom-by-atom analysis is supported by the observation that $\xi \sim {r}^{-3}$ for small $r$. 
We observed for example in case of the Rashba effect that 90\% of the Rashba strength is produced by the wave function occupying a volume in the vicinity of the nucleus given by a radius of only about 10\% (\einheit{0.25}{a.u.}) of the muffin-tin radius.\cite{Bihlmayer2006} 
We expect an analogous behavior for the DMI. 
Thus, according to Eqs.~(\ref{eq:FstO_expctval}) and (\ref{eq:SOCoperator}) also $\delta{\epsilon}^\mu_{\vcrm{k}\nu} (\vcrm{q},\hat{\vcrm{e}}_\mathrm{rot})$ is atom dependent and the DM energy is a result of atom-by-atom contributions $E_\mathrm{DMI}(\vcrm{q},\hat{\vcrm{e}}_\mathrm{rot}) = \sum_\mu E^\mu_\mathrm{DMI}(\vcrm{q},\hat{\vcrm{e}}_\mathrm{rot})$, at least in first-order perturbation theory that we discuss here throughout the paper. 
The linear fit of $D^\mu(\hat{\vcrm{e}}_\mathrm{rot}) \,q$ to $E^\mu_\mathrm{DMI}(\vcrm{q},\hat{\vcrm{e}}_\mathrm{rot})$ at the vicinity of a high symmetry point in the Brillouin zone of propagation vectors gives then the decomposition of the $\Dvec$-vector into contributions $\Dvec^{\mu}$, which satisfy
\beq
         ( \vcrm{D}^\mu\cdot\hat{\vcrm{e}}_\mathrm{rot})\, q
  \simeq \sum_{\vcrm{k}\nu}{ f(\epsilon^{0}_{\vcrm{k}\nu}, T) \,
         \delta{\epsilon}^\mu_{\vcrm{k}\nu}(\vcrm{q},\hat{\vcrm{e}}_\mrm{rot}) } \punkt
  \label{eq:EDM_mu_smearingT}
\eeq
For the interpretation of the atom dependent spiralization we refer to the discussion of the Fert-Levy model in Sec.~\ref{sec:DzyaloshinskiiVector} and in Appendix~\ref{app:sss-Fert-Levy-Model}.

Since the structure of the unit cell setup in our \abinitio calculation is inversion symmetric, the contributions of the DMI to the total energy cancel when all atoms are taken into account. 
Thus, we manually break the inversion symmetry by considering only the energy differences due to SOC from the atoms that are placed in the upper half of the unit cell.

\subsection{Magnetic anisotropy}

For the magnetic anisotropy energy the force theorem of Andersen\cite{MaAn1980} is applied, now in order to extract energy differences between collinear systems with magnetizations pointing in different directions. 
Starting point for the force theorem are self-consistent calculations including SOC, for which the magnetic moments point along the $y$ direction. 
For each system we evaluate the total energy for several directions of the magnetic moments collinearly aligned within the $xz$ plane and the $yz$ plane. 
Out of the obtained energy landscape one is able to extract $K_1$, $K_2$, and $K_3$, the principal axes of the anisotropy tensor, $\Kmatx$ (\textit{cf}.~Eq.~(\ref{eq:DvecKmatx_start})).

\section{Results and Discussion}
\label{sec:chResults}

\subsection{Spin stiffness}
\label{sec:Spinstiffness}

\begin{figure}[tb]
 \centering
   \includegraphics[width=0.45\textwidth]{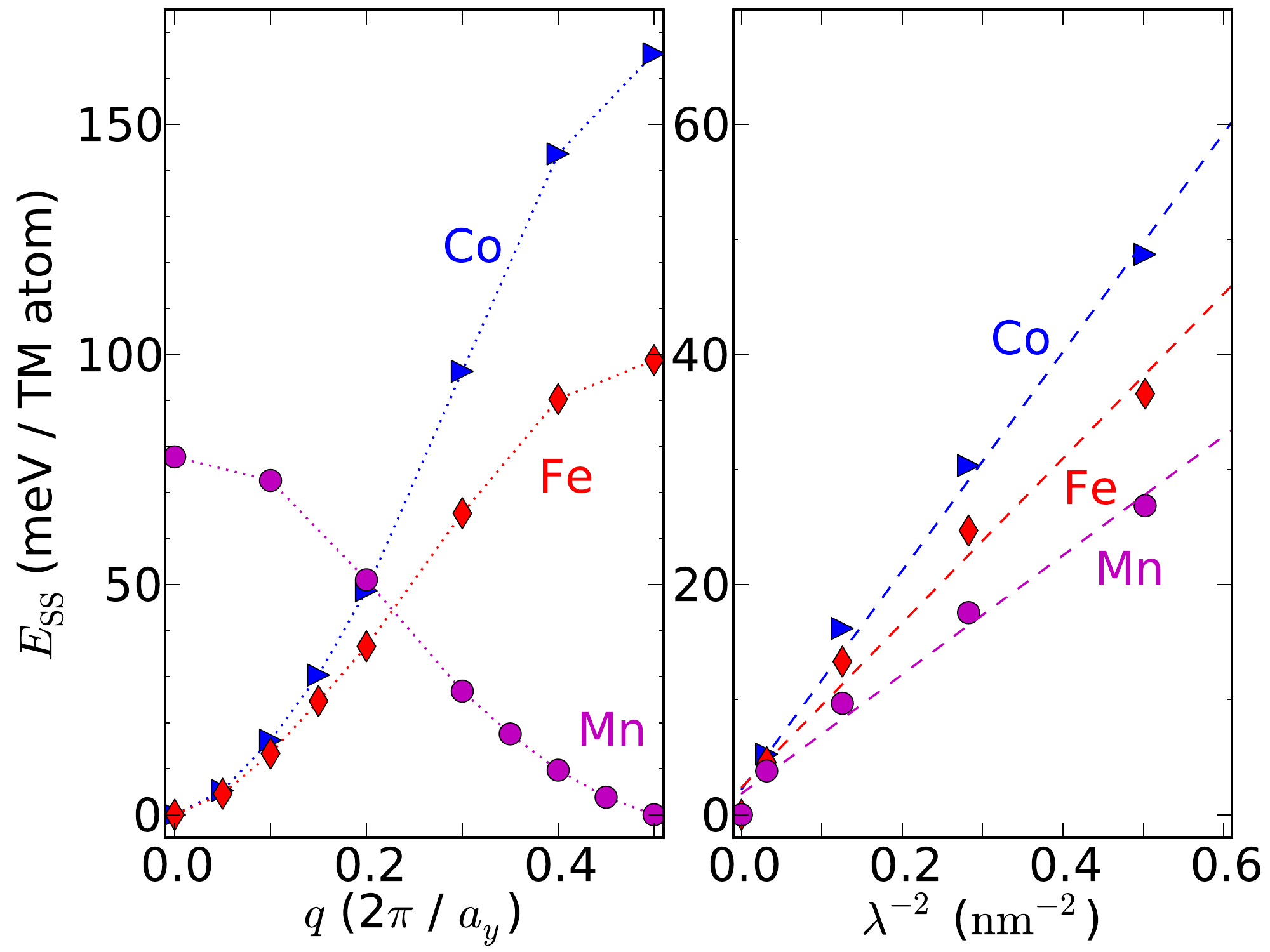}
 \caption{(color online)
          Determination of the spin stiffness: 
          In the left panel the total energies relative to their respective lowest energy, 
          $E_\mrm{SS}$, are shown as functions of the length of the wave vector, $q$, for 
          Mn, Fe, and Co chains (magenta circles, red diamonds, and blue triangles, respectively). 
          All systems show a collinear ground state, \ie, a ferromagnetic ($q=0$) ground state for the 
          Co and Fe chains and an antiferromagnetic ($q=0.5$ in units of $2\pi/a_y$) ground state for the Mn chains. 
          The right panel shows the energy as function of $\lambda^{-2}$ in the linear regime 
          with the corresponding linear fits. 
          The slope represents the spin stiffness, $\A$. 
          Note that for the Mn chain we consider the antiferromagnetic ordering vector, 
          meaning that $\lambda \rightarrow \infty$ leads to the AFM spin alignment. 
          The relative error due to the linear regression is in the order of 5\% to 8\% for the shown data range.}
 \label{fig:TM_Spst}
\end{figure}

The results of the spin-spiral energy $E_{\mrm{SS}}(q)$ for all three investigated systems as function of the wave vector $q$ along the one-dimensional Brillouin zone are summarized in Fig.~\ref{fig:TM_Spst}. 
For Co and Fe chains the minimal energy is found for the ferromagnetic state, \ie, the state with wave vector $q=0$, whereas the Mn chains align in the antiferromagnetic order. 
According to the micromagnetic model in Sec.~\ref{sec:HomInhomSpSp} (see Eq.~(\ref{eq:micmodEn2})) and the discussions in Sec.~\ref{sec:Micromagnetic Parameters} we expect in the long-wavelength limit a linear relationship between the exchange energy and the squared inverse wavelength, which is realized by these systems for a large fraction of the Brillouin zone (40\%) and shown in the right panel of Fig.~\ref{fig:TM_Spst} with the resulting fit. 
The slope gives the spin stiffness $A$. 
For the Mn system it is smallest ($\einheit{0.030}{\mrm{aJ}\mrm{nm}}$) and rises when going to Fe ($\einheit{0.041}{\mrm{aJ}\mrm{nm}}$) and Co ($\einheit{0.055}{\mrm{aJ}\mrm{nm}}$). 
The results are also collected in Table~\ref{tab:Results}. 
A small spin stiffness is favorable for the stabilization of a chiral spin spiral and in this respect the Mn chain is the most favorable system.

\subsection{The Dzyaloshinskii-vector}
\label{sec:DzyaloshinskiiVector}

\begin{figure}[tb]
 \centering
   \includegraphics[width=0.45\textwidth]{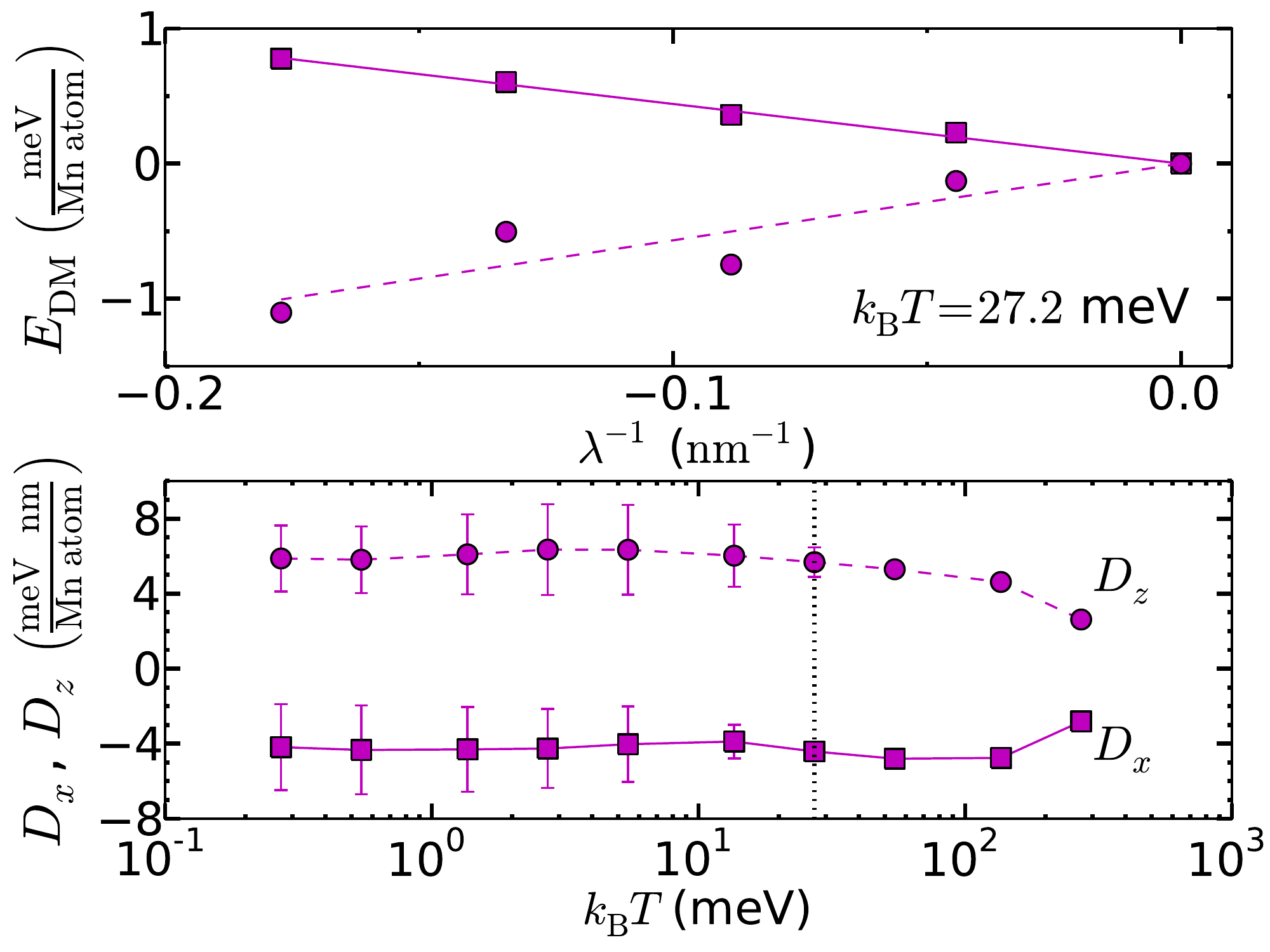}
 \caption{(color online)
          The upper panel displays for Mn/Pt(664) 
          the Dzyaloshinskii-Moriya energy $E_\mrm{DM}$ (see Eq.~(\ref{eq:EDM_smearingT})) 
          in the vicinity of the AFM state as function of the inverse wave length for flat homogeneous spirals 
          rotating in the $yz$ plane (magenta squares and solid line) and in the $xy$ plane 
          (magenta spheres and dashed line) for a temperature broadening of $k_\mrm{B}T=\einheit{27.2}{\mrm{meV}}$. 
          The slopes give the values for the components of the $\Dvec$-vector (see Eq.~(\ref{eq:DvecKmatx_start})) 
          which are shown in the lower panel as function of a broadening temperature. 
          Since for $k_\mrm{B}T=\einheit{27.2}{\mrm{meV}}$ (dotted vertical line) the values are converged 
          and the error bars are still reasonably small (for this system as well as for the other 
          two the relative error due to the linear regression is in the order of 5\% to 10\%.), the 
          corresponding values are considered in the following.}
 \label{fig:Mn_Dvec}
\end{figure}

Fig.~\ref{fig:Mn_Dvec} displays the DM energies per chain atom $E_\mrm{DM}$ and the $x$ and $z$ components of the spiralization vector $\Dvec$ for the Mn chains. 
In the upper panel we present $E_\mrm{DM}(1/\lambda, \hat{\vcrm{e}}_{x/z})$ as function of the inverse wavelength, $1/\lambda$, for clockwise rotating (negative values of $\lambda$) homogeneous spin spirals of two rotational directions $\hat{\vcrm{e}}_{x}$ and $\hat{\vcrm{e}}_{z}$. 
Analogously to the discussion of the spin stiffness, we utilize the micromagnetic model and expect a linear behavior of $E_\mrm{DM}(1/\lambda, \hat{\vcrm{e}}_{x/z}) \propto D_{x/z}\cdot1/\lambda$ for corresponding wave vectors in the vicinity of high-symmetry points in the one-dimensional Brillouin zone, $q=0$ and $q=\pi/a_y$. 
Indeed we find a linear behavior for wave vectors covering 10\% of the Brillouin zone measured from the antiferromagnetic state at $\pi/a_y$ for $E_\mrm{DM}(1/\lambda, \hat{\vcrm{e}}_{x})$. 
However, for $E_\mrm{DM}(1/\lambda, \hat{\vcrm{e}}_{z})$ we notice a periodic modulation on top of the linear behavior. 
Such oscillations can occur due to finite numerical resolutions, \eg, due to finite sampling of the Brillouin zone.\cite{ZHBB2014} 
In the lower panel of Fig.~\ref{fig:Mn_Dvec} we analyze the effect of the electronic Fermi surface broadening temperature, $T$, on the obtained slopes that correspond to the $\Dvec$-vector components and the loss of linear behavior reflected in the error bars. 
When $T$ is decreased, the values of the slopes and thus those of the $\Dvec$-vector components converge while at the same time the error bars are increasing. 
In the following, the values corresponding to $k_\mrm{B}T=\einheit{27.2}{\mrm{meV}}$ (see dotted vertical black line in the lower panel of Fig.~\ref{fig:Mn_Dvec}) are used and can be found in Table~\ref{tab:Results}. 
Among the three systems the resulting $\Dvec$-vectors show remarkable differences in direction and strength. 
Therefore, we investigate its origin in more detail in the next paragraph.

\begin{figure*}[tb]
 \centering
   \includegraphics[scale=1]{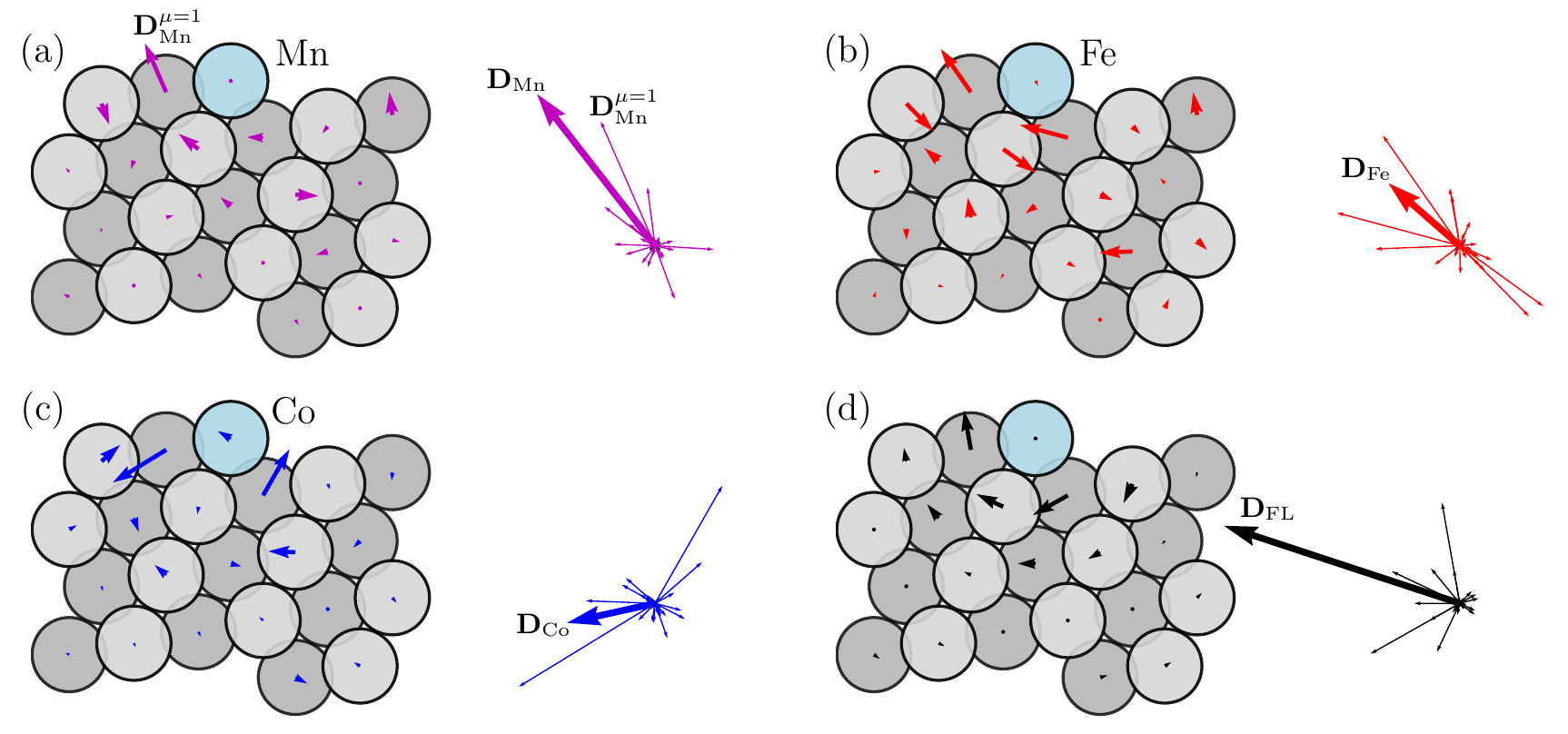}
 \caption{(color online) 
          The atom resolved contributions $\{\Dvec^\mu\}$ to the $\Dvec$-vectors are shown 
          (a) for the Mn chain, 
          (b) for the Fe chain, and 
          (c) for the Co chain, extracted from the performed \abinitio calculations, as well as 
          (d) for a TM chain when applying the Fert-Levy model (see Appendix~\ref{app:sss-Fert-Levy-Model}). 
          For each of the four cases, these contributions are depicted twice. 
          In the left image, a cross section of the step-edge structure is shown with these vectors $\{\Dvec^\mu\}$, 
          for convenience, located at the corresponding atom $\mu$ 
          (in fact, these vectors act only on the TM atoms within the chain, represented by the light blue circles). 
          In the right image, they are given with respect to the same origin. 
          In addition, the resulting $\Dvec$-vector, \ie, the sum over atoms $\mu$, $\Dvec=\sum_\mu\vcrm{D}^\mu$, 
          is printed in boldface.}
 \label{fig:TM_Dvec_AtRes}
\end{figure*}

For the three investigated systems, a more detailed study of the atom-resolved contributions to the $\Dvec$-vectors is given in Figs.~\ref{fig:TM_Dvec_AtRes}(a)-\ref{fig:TM_Dvec_AtRes}(c). 
These atom-resolved contributions, \ie, $\Dvec^\mu$ for the atom with label $\mu$, are obtained by switching on the SOC contribution for atom $\mu$ only. 
For each system they are plotted as vector with $x$ and $z$ components twice, ($i$) with respect to atom $\mu$ in the step-edge structure in the left part of each panel and ($ii$) with respect to the same origin in the right part of each panel, where in addition their sum, the $\Dvec$-vector, is shown as bold arrow. 
At first, we realize that for the Mn and Fe chain both $\Dvec$-vectors point into very similar directions. 
Although all three $\Dvec$-vectors point towards the upper step-edge, the direction of the $\Dvec$-vector of the Co chain is quite different from those of the Mn and of the Fe chain. 
The lengths of the $\Dvec$-vectors for Fe and Co chains are quite similar, but about only half as large as for the Mn chain. 
In general, the contribution $\vcrm{D}^\mu$ of the $3d$ atom itself is nearly negligible. 
The largest contributions come from atoms that are located next to the chain, albeit some contributions from some farther atoms can play a role as it is the case for Fe. 
A dominant contribution comes from the nearest-neighbor Pt atom at the upper terrace. 
For all systems they are of similar size, but for Pt next to Co, $\vcrm{D}^\mu$ points in a direction different to the Mn or Fe case (\cf~discussion at the end of this section). 
For the Co and Fe systems, each Pt atom $\mu$ with a dominant contribution $\Dvec^{\mu}$ has a vicinal atom with a $\Dvec^\mu$ vector of opposite sign and similar size. 
The dominant term for the Mn system, the nearest-neighbor atom at the upper terrace, has no counteracting contribution and consequently leads overall to a larger size of the Dzyaloshinskii-vector $\Dvec$. 
This analysis shows that, although the atom resolved contributions might have large values themselves, the sum of all contributions can still lead to a rather moderate $\Dvec$-vector due to mutual compensation.

In Fig.~\ref{fig:TM_Dvec_AtRes}(d) we show an attempt to describe the regarded structure in terms of the model proposed by Fert and Levy,\cite{FL80,LF81} where the DM energy $E_\mrm{DMI}^\mu$ is given as sum over two distinct magnetic atoms within the chain interacting with the substrate atom $\mu$ (see Appendix~\ref{app:sss-Fert-Levy-Model} for details). 
Within this model, the direction of the atom-resolved $\Dvec^\mu$-vectors is predefined to be perpendicular to the connection of the center of atom $\mu$ and the chain axis and perpendicular to the chain direction. 
This is in good agreement with the directions for the $\Dvec^\mu$-vectors for the Mn and the Fe chains (\cf~Figs.~\ref{fig:TM_Dvec_AtRes}(a) and \ref{fig:TM_Dvec_AtRes}(b) with Fig.~\ref{fig:TM_Dvec_AtRes}(d)), while it cannot be used to explain the directions for the $\Dvec^\mu$-vectors for the Co chain (\cf~Fig.~\ref{fig:TM_Dvec_AtRes}(c) with Fig.~\ref{fig:TM_Dvec_AtRes}(d)).

We next discuss the strength of the $\Dvec^\mu$-vectors. 
Applying the Fert-Levy model to a periodic infinite chain, one finds that $E_\mrm{DMI}^\mu$ vanishes in the limit $q\rightarrow0$, while its derivative and thus the $\Dvec^\mu$-vector, diverges. 
Therefore, this model is not applicable in this limit and the introduction of corrections attenuating or truncating the interaction between atoms in the infinitely long periodic chain after a certain interaction range, \eg, due to the lack of phase coherence or the presence of disorder will resolve this problem. 
Here, however, we avoid this singularity by evaluating the strength of $E_\mrm{DMI}^\mu(q_0)/q_0$ for a finite wave vector $q_0=0.05 \, \frac{2\pi}{a_y}$, which corresponds to $\lambda^{-1}=\einheit{1.77}{\mrm{nm}^{-1}}$ and thus matches in length with a $q$-vector used in the presented \abinitio calculations, see leftmost data points in upper panel of Fig.~\ref{fig:Mn_Dvec}. 
The resulting strengths of the $\Dvec^\mu$-vectors decrease with distance to the chain (see Fig.~\ref{fig:TM_Dvec_AtRes}(d)). 
The same behavior is also found for the three investigated chains, albeit the length of the vectors cannot be explained by the distance to the chain only. 
In Appendix~\ref{app:sss-Fert-Levy-Model} we furthermore show that the strength decays with distance much faster when the magnetic moments of the atoms within the chain show a AFM short-range order, as compared to a FM short range order in the same chain. 
This observation, however, cannot be extracted from the \abinitio results, \eg, when comparing the Mn chain (see Fig.~\ref{fig:TM_Dvec_AtRes}(a)) to the Fe or the Co chain (see Figs.~\ref{fig:TM_Dvec_AtRes}(b) and \ref{fig:TM_Dvec_AtRes}(c)). 
In conclusion, with regard to the structure of an infinite chain of magnetic atoms, we find that the model of Fert and Levy does not capture the diverse behavior of the three considered chains and we advise the application of this model to chains with some precaution. 
A more thorough investigation of the predictive power of the Fert-Levy model with respect to films and heterostructures would be interesting.

Finally, we provide some arguments why the directions of $\vcrm{D}^\mu$ from Pt atoms next to Co contributing to the total DMI vector are so different as compared to those next to Mn or Fe. 
From a simple tight-binding model that we developed in Ref.~\onlinecite{KSZ+2014}, we identified spin-flip transitions between occupied and unoccupied states as the relevant process for a non-vanishing DMI. 
For the Mn chain, the spin-up (spin-down) channels are entirely occupied (unoccupied) and all transitions yield a contribution to the DMI. 
Going now to Co, some spin-down states become occupied and transitions into these states do not contribute anymore to the DMI. 
Since the remaining empty states exhibit particular orbital characters, the $\vcrm{D}^\mu$ vector may well be rotated as compared to Mn. 
The situation for Fe is similar to Mn: most of the spin-down states are still unoccupied. 
Of course, a quantitative analyze requires many more details, such as bandwidths, the nature of the chemical bond etc., but this goes beyond the scope of this paper.

\subsection{The anisotropy tensor}
\label{sec:results.anisotropy}

\begin{figure}[tb]
 \centering
   \includegraphics[scale=1]{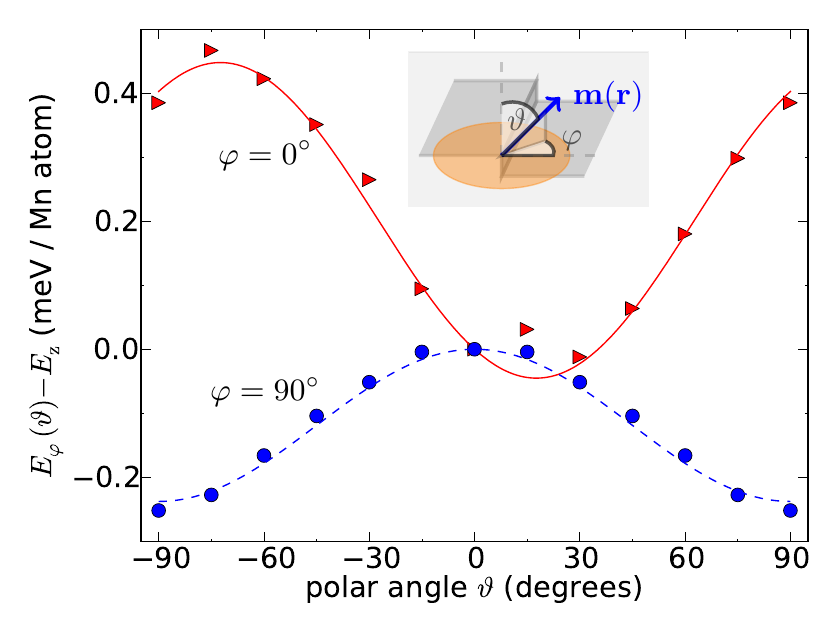}
 \caption{(color online)
          Magnetic anisotropy energy for Mn/Pt(664). 
          The energy is plotted for magnetic moments pointing along directions 
          discretized by the angles $\varphi$ and $\vartheta$ relative to the 
          orientation in $z$ direction. 
          The symbols represent \abinitio calculated energy differences, 
          whereas the fit functions correspond to Eq.~(\ref{eq:fitMAE}). 
          In the case of $\varphi=0^\circ$ (solid red line) the $xz$ plane and 
          in the case of $\varphi=90^\circ$ (dashed blue line) the $yz$ plane is sampled.}
 \label{fig:Mn_AniE}
\end{figure}

Following the findings of Sec.~\ref{sec:Spinstiffness} we investigate the magnetic anisotropy tensor for the ferromagnetic order for the Co and Fe chains, and for the antiferromagnetic order for the Mn chains. 
The two required data sets for the latter system are shown in Fig.~\ref{fig:Mn_AniE}. 
The fit functions represent the leading order term and have the form
\beq
  E_\varphi(\vartheta) = A_\varphi \cdot \cos^2\left(\vartheta + B_\varphi\right) - E_z
  \label{eq:fitMAE}
\eeq
with the energy offset $E_z=E_{\varphi=0}(\vartheta=0)$, the polar angle $\vartheta$ as argument, the azimuth angle $\varphi\in\{ 0^\circ , 90^\circ \}$ as parameter (see the inset in Fig.~\ref{fig:Mn_AniE}), and fit parameters $A_\varphi$ and $B_\varphi$. 
The mirror plane perpendicular to the chain direction is reflected by the fact that $B_{\varphi=90^\circ}=0$, leading to a symmetric function with respect to $\vartheta$. 
The resulting hard, medium, and easy axes for all three systems are summarized in Table~\ref{tab:Results}. 
In the following we use the easy axis as energy offset.

With respect to the resulting principal components the Mn system appears to be the most promising candidate for a non-collinear ground state. 
The anisotropy energies for the medium and the hard axis are the smallest compared to those of the other two systems. 
In addition, the easy axis points along the chain direction which is of relevance for the following reason: 
The only spin-orbit driven spin spiral that the $\Dvec$-vector (perpendicular to the chain axis) can stabilize are of cycloidal character, meaning that the spiral rotation plane always contains the direction along the chain. 
Thus, the rotation over the easy axis is achieved automatically, regardless of the rotation axis. 
For the Co chains the easy axis is at about $61^\circ$ tilted towards the upper terrace, which is in satisfying agreement with other experimental\cite{GDM+2002} and theoretical findings.\cite{KEDF2002,ShMP2004,ULS+2004} The easy axis of the system containing the Fe chains is directed in approximately the same direction as the one for the Co chains. 
A remarkable finding for the Fe system is the strength as well as the orientation of the hard axis. 
It not only exhibits the largest value among all three systems, but also is oriented along the chain direction. 
Therefore, it shows the most unfavorable setup for a cycloidal spiral to appear since a rotation over the hard axis would be unavoidable.

\subsection{Magnetic ground states}

\begin{figure*}[tb]
 \centering
   \includegraphics[scale=1]{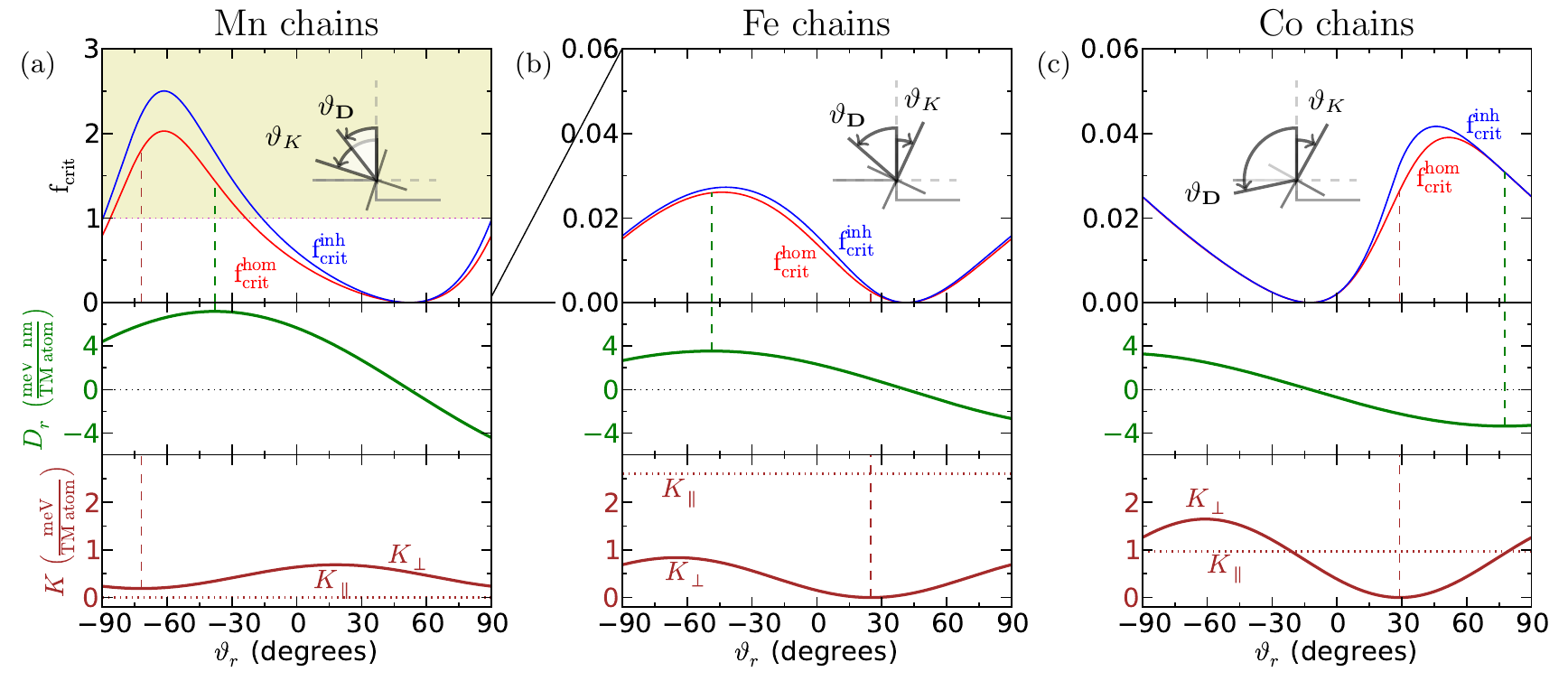}
 \caption{(color online)
          In the upper panels we show the values of the functions 
          $\mrm{f}_\mrm{crit}^\mrm{hom}$ and $\mrm{f}_\mrm{crit}^\mrm{inh}$
          for the appearance of homogeneous and inhomogeneous spin spirals 
          (\cf~Eqs.~(\ref{eq:Crit_hss}) and (\ref{eq:Crit_iss}), respectively) in 
          (a) Mn, (b) Fe, and (c) Co chains as functions of $\vartheta_r$, the 
          direction of the rotation axis, see Fig.~\ref{fig:mod_stepedge}. 
          For each system the inset shows the relative orientation of the 
          $\Dvec$-vector with respect to the principal axes of the anisotropy tensor. 
          In the lower panels the corresponding parameters $D_r$, $K_\parallel$, and $K_\perp$ are plotted. 
          In the last two systems the critical threshold of $1$ is missed by more than one order of magnitude. 
          For the Mn chains, however, both criteria are fulfilled, 
          and their respective curves reach their maxima for $\vartheta_r=-62^\circ$. 
          This can be seen as a compromise between 
          finding the largest DMI contribution (dashed green line) and having the 
          smallest anisotropy energy $K_\perp$ (dashed brown line).}
 \label{fig:TM_result}
\end{figure*}

\begin{table*}[bt]
 \begin{center}
 \begin{ruledtabular}
 \begin{tabular}[c]{ccccccccccc}
    TM & $\begin{matrix} \A                    \\ \left(\frac{\mrm{meV\;nm}^2}{\mrm{TM\;atom}}\right) \end{matrix}$
       & $\begin{matrix} \vert\Dvec\vert       \\ \left(\frac{\mrm{meV\;nm}}{\mrm{TM\;atom}}\right) \end{matrix}$
       & $\begin{matrix} \vartheta_{\Dvec}     \\ (\mrm{degrees}) \end{matrix}$
       & $\begin{matrix} K_1                   \\ \left(\frac{\mrm{meV}}{\mrm{TM\;atom}}\right) \end{matrix}$
       & $\begin{matrix} K_2                   \\ \left(\frac{\mrm{meV}}{\mrm{TM\;atom}}\right) \end{matrix}$
       & $\begin{matrix} K_3                   \\ \left(\frac{\mrm{meV}}{\mrm{TM\;atom}}\right) \end{matrix}$
       & $\begin{matrix} \vartheta_{K}         \\ (\mrm{degrees}) \end{matrix}$
       & $\begin{matrix} \vartheta_r^\mrm{max} \\ (\mrm{degrees}) \end{matrix}$
       & $\begin{matrix} \mu_\mrm{mag}         \\ (\mu_\mrm{B}) \end{matrix}$
       & $\begin{matrix} \mu_\mrm{orb}         \\ (\mu_\mrm{B}) \end{matrix}$ \\
    \hline
    Mn  & $ 52$  &  $ 7.2$   & \;\;$ -38$  & $0.69$  & $0.0 $  & $0.19$ & $-72$  &  $-62$  & $4.00$ & $0.04$ \\
    Fe  & $ 72$  &  $ 3.5$   & \;\;$ -49$  & $0.84$  & $2.61$  & $0.0$  &  $25$  & $(-42)$ & $3.27$ & $0.14$ \\
    Co  & $ 97$  &  $ 3.4$   &     $-102$  & $1.65$  & $0.97$  & $0.0$  &  $29$  & $(+46)$ & $2.20$ & $0.19$ \\
 \end{tabular}
 \end{ruledtabular}
 \caption{Collected results for the three investigated TM chains on Pt(664) step-edges: 
          Spin stiffness $\A$, 
          absolute value of the Dzyaloshinskii-vector $\vert\Dvec\vert$ 
          and its orientation $\vartheta_{\Dvec}$, 
          principal components of the anisotropy tensor $\Kmatx$ 
          (\cf~Eq.~(\ref{eq:DvecKmatx_start})), $K_1$, $K_2$, $K_3$, 
          and $\vartheta_{K}$, the orientation of the principal axis corresponding to $K_3$, 
          the rotation angle $\vartheta_r^\mrm{max}$, 
          as well as the spin magnetic moment $\mu_\mrm{mag}$ and orbital magnetic moment $\mu_\mrm{orb}$ 
          of the TM atom for the case that the spin-quantization axis points along the easy axis. 
          All angles are measured with respect to the $z$-axis, see Fig.~\ref{fig:mod_stepedge} and the insets of 
          Figs.~\ref{fig:TM_result}(a), \ref{fig:TM_result}(b), and \ref{fig:TM_result}(c). 
          $\vartheta_r^\mrm{max}$ indicates where the function $\mrm{f}_\mrm{crit}^\mrm{inh}$ gets maximal, 
          representing the planar inhomogeneous spin-spiral of lowest energy among all spirals. 
          Only for the Mn chains this energy is lower than the one for the collinear state 
          (\ie, criterion (\ref{eq:Crit_iss}) is satisfied) 
          and a spin-spiral state is formed as ground state. 
          For the Fe and Co chains the collinear state always remains lower in energy. 
          Note that $K_2$, the anisotropy along the chain, is the easy axis for the Mn chains 
          but the hard axis for the Fe system. 
          $\A$, $\vert\Dvec\vert$, and $K_i$ can be expressed in units directly 
          compatible to the micromagnetic equation~(\ref{eq:micmodEn}) dividing the parameter values by 
          $\Delta=\einheit{0.282}{\text{nm / TM atom}}$.}
 \label{tab:Results}
 \end{center}
\end{table*}

In the previous Secs.~\ref{sec:Spinstiffness}, \ref{sec:DzyaloshinskiiVector}, and \ref{sec:results.anisotropy} we extracted the parameters that now can be used to evaluate the criteria for the appearance of homogeneous and inhomogeneous spin spirals (see Eqs.~(\ref{eq:Crit_hss}) and (\ref{eq:Crit_iss}), respectively) and their respective properties.\footnote{In Sec.~\ref{sec:HomInhomSpSp} it was pointed out that for the same set of parameters the criterion for the appearance of an inhomogeneous spin spiral is always more likely to be fulfilled than the one for the homogeneous spiral. 
Nevertheless we will still present our findings regarding both spiral types, which enables the reader to compare.} 
Those criteria depend on the spin stiffness $\A$ as well as $D_r$, the projection of the Dzyaloshinskii-vector onto the rotation axis $\hat{\vcrm{e}}_\mrm{rot}$, and $K_\parallel$, and $K_\perp$, the two principal axes of the anisotropy tensor that describe spins rotating in the plane perpendicular to the rotation axis (see Fig.~\ref{fig:mod_stepedge}). 
Evaluating Eqs.~(\ref{eq:Crit_hss}) and (\ref{eq:Crit_iss}) the resulting magnetic ground state is determined by the functions $\mrm{f}_\mrm{crit}^\mrm{hom}(\vartheta_r)$ and $\mrm{f}_\mrm{crit}^\mrm{inh}(\vartheta_r)$ and whether their value exceeds the critical threshold of 1 for at least one rotation direction, described by the rotation angle $\vartheta_r$. 
As we can see in Fig.~\ref{fig:TM_result} the Mn chains indeed fulfill both criteria when the direction of the spin-rotation axis, around which the magnetic moments of the spiral rotate, is in the regime between about $-90^\circ \leq \vartheta_r \leq -20^\circ$. 
The maximum values of $\mrm{f}_\mrm{crit}^\mrm{hom}(\vartheta_r)$ and $\mrm{f}_\mrm{crit}^\mrm{inh}(\vartheta_r)$ are obtained for $\vartheta_r=-62^\circ$, which at the same time represents the minimum of the total energy, \ie, the magnetic ground state. 
This rotation angle can be understood as a compromise between the optimal DMI contribution ($\vartheta_r=-38^\circ$, rotation axis parallel to $\Dvec$-vector) and the minimal MAE barriers ($\vartheta_r=-72^\circ$, rotation axis along $K_3$, the hard axis). 
Since $D_r$ is positive for $\vartheta_r=-62^\circ$, the obtained magnetic structure is a left-rotating spiral, which modulates the otherwise antiferromagnetic order. 
As the magnetic anisotropy within the $yz$ plane (see Fig.~\ref{fig:Mn_AniE}, dashed blue curve) is small, the findings for homogeneous and inhomogeneous spirals are quite similar. 
For the same set of parameters, Eqs.~(\ref{eq:enehom}) and (\ref{eq:laminh}) lead to a spiral length of $\lambda=\einheit{15.7}{\mrm{nm}}$ for a homogeneous spin spiral and $\lambda=\einheit{16.3}{\mrm{nm}}$ for an inhomogeneous spiral. 
Both values correspond to a period length of about 60 atoms along the chain, which is equivalent to an average rotation angle of $\varphi\approx174^\circ$ between neighboring Mn atoms and a $\vcrm{q}$-vector of $\vcrm{q}=-\frac{1}{60} \, \frac{2\pi}{a_y} \, \hat{\vcrm{e}}_y$, measured from the AFM alignment. 
These large period lengths justify in retrospect our ansatz of a micromagnetic model. 
Employing Eqs.~(\ref{eq:enehom}) and (\ref{eq:eneinh}) we find an averaged energy gain of $\Delta E=\einheit{-0.106}{\mrm{meV}}$ per chain atom ($\Delta E=\einheit{-0.376}{\mrm{meV\;nm^{-1}}}$) for the homogeneous spin spiral and $\Delta E=\einheit{-0.113}{\mrm{meV}}$ per chain atom ($\Delta E=\einheit{-0.399}{\mrm{meV\;nm^{-1}}}$) in the case of an inhomogeneous spin spiral.

In contrast to the analysis of the Mn chains, the obtained parameters for the Fe and Co chains confirm a ferromagnetic ground state, which is in line with previous studies.\cite{BRBB2006,BBBR2006,HonolkaPRL09} 
For one part the resulting DMI is not large enough to change the collinear order favored by the spin stiffness, which we trace back to oppositely directed atom-resolved contributions to the $\Dvec$-vector of the Pt atoms nearby the chain (\cf~Figs.~\ref{fig:TM_Dvec_AtRes}(b) and \ref{fig:TM_Dvec_AtRes}(c)). 
On the other hand, the magnetic anisotropy causes energy barriers that prevent the system from forming a non-collinear ground state. 
Especially for the case of Fe chains the formation of a spin spiral turns out to be energetically unfavorable, as the spin moments would have to rotate over the hard axis, as mentioned in Sec.~\ref{sec:results.anisotropy}.

We conclude the investigation on the magnetic ground state with a brief discussion on the possibility of finding non-planar spin spirals. 
For systems with orthorhombic anisotropy, phase diagrams are known\cite{HeBB2011} that take such three-dimensional non-collinear spin structures into account. 
However, to make our parameters match the Ansatz made in Ref.~\onlinecite{HeBB2011} one has to assume that the $\Dvec$-vector is oriented along one of the two principal axes of the anisotropy vector, $K_1$ or $K_3$. 
This is to some extent only reasonable for the Fe chains where the angle between easy axis direction and $\Dvec$-vector is $16^\circ$ (see insets in Fig.~\ref{fig:TM_result}). 
In addition this system is the best candidate for a three-dimensional spiral since a rotation over the unfavorable hard axis is avoided, so that we restrict our analysis onto the Fe chains only. 
Following the notation of Ref.~\onlinecite{HeBB2011} we arrive at $D_I=0.32$ and $K_I=-0.48$, when the $\Dvec$-vector is assumed to point along the easy axis direction. 
Thus, we miss the critical regime of $D_I>1$ by a factor of 3, and this pair of parameters distinctly lies in the collinear region (\cf~Fig.~3(b) in Ref.~\onlinecite{HeBB2011}). 

\subsection{Formation of domain walls}

Although for Fe and Co chains the DMI is not strong enough to introduce a chiral magnetic ground state its presence can influence the formation of domain walls.\cite{HeBB2008} 
We follow the analysis of chiral domain walls put forward by Dzyaloshinskii,\cite{Dzya1965} but apply this analysis to the ferromagnetic Fe and Co chains. 
Once again the starting point is the energy functional as given in Eq.~(\ref{eq:micmodEn}), now with the boundary condition
\beq
  \vcrm{m}(y) \xrightarrow{y \rightarrow \pm \infty} \pm \vcrm{m}_\mrm{easy} \punkt
  \label{eq:boundaryWall}
\eeq
By this constraint a rotation by $180^\circ$ is forced to take place spreading within the infinite chain. 
A distinction is made between a Bloch wall (helical rotation) and a N\'eel wall (cycloidal rotation). 
For both cases a characteristic width of the planar domain wall can be defined by\cite{HeBB2008}
\beq
  w = \frac{1}{\pi}\cdot\sqrt{\frac{\A}{K}} \komma
\eeq
where $K$ represents the anisotropy energy for magnetic moments that point perpendicular to the easy axis direction within the spin rotation plane. 
The expression for the minimal energy reads\cite{HeBB2008}
\beq
  E = \frac{2}{\pi}\cdot\sqrt{\A K} - \frac{\vert \vcrm{\hat{e}}_\mrm{rot}\cdot\Dvec \vert}{2} \komma
  \label{eq:energ_domain}
\eeq
where for the N\'eel wall $\vcrm{\hat{e}}_\mrm{rot}\cdot\Dvec$ is equal to the expression in Eq.~(\ref{eq:Dr}), but vanishes for a Bloch wall ($\vcrm{\hat{e}}_\mrm{rot} \perp \Dvec$). 
Thus, only for a N\'eel wall a preference in the rotation direction is expected and N\'eel walls can be realized even if the MAE favors a Bloch wall. 
Note, that the energetically favored rotational sense of the domain wall is accounted for by the minus sign and the absolute value of the second term in Eq.~(\ref{eq:energ_domain}).

The resulting domain wall energies as well as the predicted wall widths for Fe and Co chains are listed in Table~\ref{tab:domainwalls}. 
Since for both systems the easy-axis direction is perpendicular to the chain direction, the rotation axis is fixed by Eq.~(\ref{eq:boundaryWall}) and the chain direction. 
If the easy axis points along the chain direction, $\vcrm{\hat{e}}_\mrm{rot}$ is a compromise between magnetic anisotropy and DMI as it was the case for the ground-state analysis. 
In such a case no Bloch wall can be established.

For the Fe chains, a Bloch wall is energetically more favorable than the N\'eel wall even when the DMI contribution is taken into account, so that we do not expect a preference in the rotational sense for the domain walls for this system. 
One reason is that a N\'eel wall forces a rotation of the spins over the chain direction that is the hard axis of this system. 
Furthermore, the rotation plane is predefined by the easy axis direction. 
Since the $\Dvec$-vector is oriented nearly within this plane, the projection to the rotation axis $\vcrm{\hat{e}}_\mrm{rot}$ is relatively small.

For the Co chains, the energy of the Bloch wall, $E_\mrm{B}$, is already by more than \einheit{6}{meV} higher in energy than the corresponding N\'eel wall, $E_\mrm{N}^\mrm{noDMI}$, where the DMI contribution is neglected. 
When in Eq.~(\ref{eq:energ_domain}) the DMI contribution is taken into account the preference of a N\'eel wall is even higher. 
This gain in energy is achieved only for a right-rotating domain wall. 
This is because the rotation angle $\vartheta_r=29^\circ$ that describes a rotation plane perpendicular to the easy axis leads to a negative $D_r$, see Fig.~\ref{fig:TM_result}(c). 
Such a spin-orbit driven preference of a particular rotational sense of the N\'eel wall should be observable in an experiment.

\begin{table}[bt]
 \begin{center}
 \begin{ruledtabular}
 \begin{tabular}[c]{ccccc}
    TM & $\begin{matrix} E_\mrm{B} \\ \left(\mbox{meV}\right) \end{matrix}$ &
         $\begin{matrix} w_\mrm{B} \\ \left(\mrm{nm}\right) \end{matrix}$ &
         $\begin{matrix} E_\mrm{N} \ (E_\mrm{N}^\mrm{noDMI}) \\ \left(\mbox{meV}\right) \end{matrix}$ &
         $\begin{matrix} w_\mrm{N} \\ \left(\mrm{nm}\right) \end{matrix}$ \\
    \hline
    Fe & 17.49 & 2.94 & 29.06 (30.82) & 1.67 \\
    Co & 28.26 & 2.42 & 17.73 (21.68) & 3.15 \\
 \end{tabular}
 \end{ruledtabular}
 \caption{The domain wall energies for a Bloch wall and a N\'eel wall, $E_\mrm{B}$ and $E_\mrm{N}$ respectively, 
          as well as the corresponding wall widths, $w_\mrm{B}$ and $w_\mrm{N}$, are listed. 
          Due to the DMI, the N\'eel wall always exhibits a certain rotational sense that lowers 
          the energy with respect to its value without taking the DMI into account ($E_\mrm{N}^\mrm{noDMI}$). 
          For the Fe chains the Bloch wall is energetically always more favorable, 
          for the Co chains the N\'eel wall is preferred.}
 \label{tab:domainwalls}
 \end{center}
\end{table}

\section{Summary and outlook}
\label{sec:chSummary}

\begin{figure}[tb]
 \centering
   \includegraphics[trim = 0mm 0mm 54mm 0mm, clip, width=0.45\textwidth]{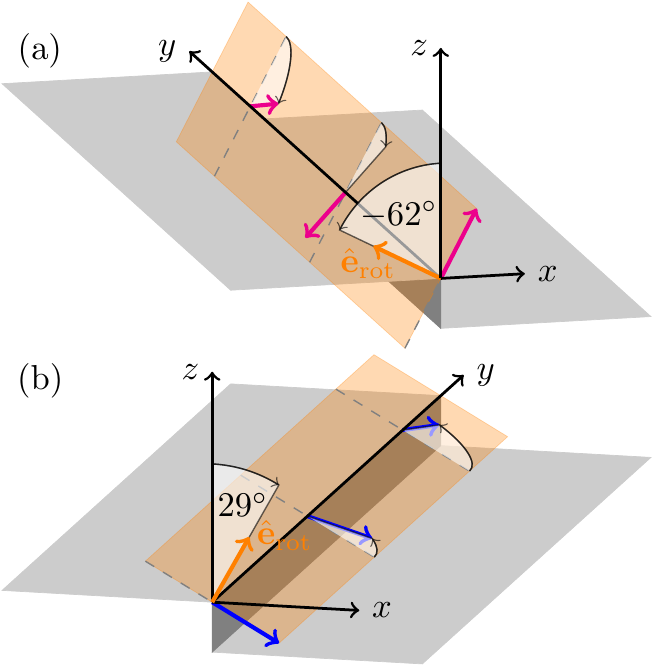}
 \caption{(color online)
          Schematic visualization of the energetically preferred rotational sense for 
          (a) the ground-state of the Mn chain (left-rotating spin-spiral) and 
          (b) the domain wall for the Co chain (right-rotating N\'eel wall).}
 \label{fig:Summary}
\end{figure}

Density-functional theory (DFT) calculations were performed to study the magnetic interactions in Mn, Fe, and Co chains at Pt step-edges. 
These calculations allow to extract parameters for a micromagnetic model that takes into account the spin-stiffness constant, $\A$, the magnetic anisotropy tensor, $\Kmatx$, and the $\Dvec$-vector, which arises from the Dzyaloshinskii-Moriya interaction (DMI). 
Using this model, the magnetic ground state for the three investigated systems is determined employing two different instability criteria for the appearance of spin-orbit driven non-collinear structures, one for the homogeneous and one for the inhomogeneous spin spiral. 
The main results are listed in Table \ref{tab:summary}.

Our results predict a spiral magnetic ground state for the Mn chains, that modulates the antiferromagnetic order with a period length of about $\einheit{16}{\mrm{nm}}$ or 50-60 atoms along the chain. 
These findings establish Mn as a promising candidate for experimental research groups to investigate the DMI in 1D systems. 
A new aspect of this system, different to the systems studied in the literature, is the non-trivial direction of the $\Dvec$-vector, that is not fully determined by symmetry. 
As a result the spiral rotates in a plane that is tilted by about $62^\circ$ towards the upper terrace (see Fig.~\ref{fig:Summary}(a)). 
For the Fe and Co chains we conclude that the formation of a non-collinear spiral magnetic structure is unlikely. 
For one part, this is due to magnetic anisotropies that are larger compared to the Mn chains. 
For the other part, their $\Dvec$-vectors are too small to overcome these anisotropy barriers. 
A detailed atom-resolved analysis of this quantity showed that their moderate strengths are due to compensation of the atomic contributions. 
For Co, the results are consistent with recent findings for the Co zigzag chain on Ir(001) (5$\times$1),\cite{DupeNJP2015} for which also no spiraling solution was observed. 
On the other hand, the Fe/Pt step-edge behaves different to the biatomic Fe chain Ir(001),\cite{MMW+2012} for which a spiral with a short period pitch was observed.

The calculated directions and strengths of the $\Dvec$-vectors for the different chains were compared to those that result from the model of Fert and Levy. 
It appears that this model reproduces to some extent the directions of the $\Dvec$-vectors of the Mn and the Fe chains. 
For the Co chain, however, it fails to describe the direction of $\Dvec$ correctly. 
We noticed that the model of Fert and Levy may be used with some precaution at least for one-dimensional chains as the micromagnetic DM vector diverges for infinite chains in the limit of long wavelengths. 
A more thorough investigation of the predictive power of the Fert-Levy model with respect to films and heterostructures would be interesting.

Furthermore, an analysis of planar domain wall structures for the Fe and the Co chains was presented. 
It appears that in the Fe system a Bloch wall is energetically more favorable. 
Since this type of domain wall is by symmetry not affected by the DMI, a preferred sense in the rotation direction is not expected. 
In contrast, the Co chains form a N\'eel wall, which shows a homochiral preference in the wall rotation that is caused by the DMI (see Fig.~\ref{fig:Summary}(b)).

We encourage experimental groups to verify our findings for the Mn chains in terms of the magnetic ground state. 
Furthermore, a statistically preferred rotational sense of domain walls in the Co chains should be observable in experiment. 
This could add a substantial aspect to the understanding of magnetism in low-dimensional systems and could provide some insight into the consequences of surface and interface roughness on the DMI. 
Previous investigations revealed a strong dependence of the MAE on the number of transition-metal strands in the chain  \cite{GDM+2004,BBBR2006,HLK+2009}. 
For example, a strong softening of the MAE was observed for Fe double-chains \cite{HLK+2009}, \ie, magnetic parameters may be tuned as functions of the number of strands to meet the criterion for a chiral ground state.

In this paper we focused exclusively on infinite periodic chains. 
Here we comment briefly on the magnetism for chains of finite lengths. 
We may discuss the finiteness in terms of a boundary effect, which are strongest where the chain terminates and whose effects decay away from the boundary into the chain. 
This affects shorter chains stronger than larger chains. 
Thus, in the center of larger chains we expect the same behavior as for periodic chains. 
In general, due to the finiteness of the chain three additional factors may play a role: 
(i)   Atoms in a finite chain lose the mirror symmetry in a plane normal to the chain direction. 
(ii)  Thus, edge effects of finite chains result in non-vanishing components of $\Dvec$-vectors along the chain direction. 
Although the remaining non-vanishing component is small when averaged across the finite chain, locally we expect an additional tilt of the magnetic moments and subsequently an additional energy gain. 
This supports the formation of a chiral magnetic ground state in a finite chain over an infinite one, similar to the surface twist in films of B20 alloys that stabilize the skyrmions phase in films over B20 bulk alloys.\cite{RybakovPRB2013} 
Even if we assume that the electronic structure at the boundary of the finite chain does not change and all microscopic magnetic parameters remain unchanged, the micromagnetic DMI experience a change due to symmetry and this additional tilt of the magnetization  has been investigated by S. Rohart and A. Thiaville\cite{RohartThiavillePRB2013} but not for chains but for nanostructures. 
(iii) The change of the electronic structure at the boundary of the chain is an additional factor. 
Actually we investigated this for finite clusters\cite{Bauer_unpublished} and it might be an important effect. 
Then all micromagnetic parameters change, but most affected are the DMI and MAE. 
This can modify the threshold for the occurrence of a chiral magnetism in the chain. 
If the chain length becomes below two times the length scale, where the electronic structure is modified due to the presence of the finiteness of the chain, nothing can be said about the magnetic property of the short chain. 
Additional \abinitio studies are required.

\begin{table}[bt]
 \begin{center}
  \begin{ruledtabular}
   \begin{tabular}[c]{ccccc}
     TM  & GS (no SOC) & GS (with SOC) & easy axis         & DW type \\
     \hline
     Mn  &       AFM   &    $\ell$-SS  & $\parallel$ chain & ---     \\
     Fe  &        FM   &           FM  & $\perp$ chain     & BW      \\
     Co  &        FM   &           FM  & $\perp$ chain     & r-NW
   \end{tabular}
  \end{ruledtabular}
  \caption{Summary of the outcome of the paper: 
           Whereas in the absence of SOC only collinear ground-states (GS) occur (FM and AFM), 
           the Mn chain forms a left-rotating homochiral spin spiral ($\ell$-SS) when SOC is taken into account. 
           The easy axis can point along the chain (Mn) or perpendicular to it (Fe and Co). 
           For the definition of $\vartheta_{K}$ see Fig.~\ref{fig:TM_result} or Eq.~(\ref{eq:theta_K}). 
           The analysis of the domain wall (DM) type reveals that the Fe chain prefers a Bloch wall (BW) 
           whereas for the Co chain a right-rotating N\'eel wall (r-NW) is energetically favored.}
  \label{tab:summary}
 \end{center}
\end{table}

On the methodological side we showed that for a spin-lattice model of classical spins on a Bravais lattice including Heisenberg and Dzyaloshinskii-Moriya interaction the homogeneous spin-spiral is an exact solution if the rotation vector of the spin-spiral points either parallel or antiparallel to the $\Dvec$-vector, representing a solution of two different chiralities. 
This has important consequences since the spin-spiral state is a state that is frequently employed in the first-principles context using density functional theory. 
One consequence is that the slope and the curvature of the spiral energy as function of the wave vector as calculated in density functional theory provides directly the spin stiffness and the spiralization that enter a material specific micromagnetic model.

\section{Acknowledgment}

We would like to thank Albert Fert, Miriam Hinzen, Daniel A.\ Kl\"uppelberg and Christoph Melcher for fruitful suggestions and stimulating discussions during the course of this work. 
We gratefully acknowledge computing time on the JUROPA supercomputer provided by the J\"{u}lich Supercomputing Centre (JSC). 
B.S.\ acknowledges funding by the HGF-YIG Programme VH-NG-717 (Functional Nanoscale Structure and Probe Simulation Laboratory, Funsilab). 
B.Z.\ and S.B.\ acknowledge funding from the European Union's Horizon 2020 research and innovation programme under grant agreement No 665095 (MAGicSky).

\appendix

\section{Relation between micromagnetic and spin-lattice model}
\label{app:micro-model}

A natural starting point for a multiscale analysis of a complex magnetic structure is the spin-lattice Hamiltonian
\beq
    E\{\vcrm{S}\}
  =   \sum_{i<j} {\left[   J_{ij}~\vcrm{S}_{i} \cdot \vcrm{S}_{j}
                         + \vcrm{D}_{ij} \cdot \left(\vcrm{S}_{i} \times \vcrm{S}_{j} \right) \right] }
    + \sum_{i}\vcrm{S}_{i}^\mathrm{T} \, \Kmatx_{i} \, \vcrm{S}_{i} \komma
  \label{eq:spin-model}
\eeq
where $J_{ij}$ is the exchange integral between atoms at sites $i$ and $j$, $\vcrm{D}_{ij}$ is the Dzyaloshinskii vector and $\Kmatx_{i}$ is the microscopic on-site anisotropy tensor. 
If these parameters are determined from first principles, one refers to a realistic spin-lattice model. 
Assuming lattice periodicity it follows that $J_{ij}= J_{|j-i|}$, $\vcrm{D}_{ij} = \vcrm{D}_{j-i}=-\vcrm{D}_{i-j}$, and $\Kmatx_i = \Kmatx_0$ for all sites $i$, and considering that the exchange interaction and the DMI are even and odd functions, respectively, with respect to inversion symmetry. 
In this appendix we relate these parameters to the micromagnetic parameters of model~(\ref{eq:micmodEn}).

If we assume that the magnetic structure is slowly varying along the chain, meaning that the magnetic moments rotate on a length scale that is much larger than the interatomic distance, then it is certainly possible to choose for the magnetization direction a continuous normalized function $\vcrm{m}(y)$ with $|\vcrm{m}(y)|=1$, such that $\vcrm{m}(j\,\Delta) = \vcrm{S}_j$, where $\Delta$ denotes the spacing between the lattice points along the $y$-axis. 
If we assume further that $\vcrm{m}$ does not vary much on a length scale at which the interactions $J$ and $\vcrm{D}$ are relevant, then the interactions can be considered local over that length scale which is consistent with the formulation of the interactions in the micromagnetic energy functional~(\ref{eq:micmodEn}). 
Under these conditions one can Taylor expand $\vcrm{S}_{j}=\vcrm{m}(j\Delta)$ around $\vcrm{m}(i\,\Delta)$. 
The energy expression~(\ref{eq:spin-model}) treated within the lowest relevant order reads then 
\begin{widetext}
\beq
  E = \sum_i\Delta
      \left[   \sum_{j>i}
               \left[ -\frac{1}{2}(j-i)^2 \Delta J_{|j-i|}~\dot{\vcrm{m}}^2(i\Delta)
                      +(j-i)\vcrm{D}_{j-i} \cdot \left(\vcrm{m}(i\Delta) \times \dot{\vcrm{m}}(i\Delta) \right)\right]
             + \frac{1}{\Delta}\vcrm{m}(i\Delta)^\mathrm{T} \, \Kmatx_0 \, \vcrm{m}(i\Delta) \right] \komma
  \label{eq:spin-model_taylor-expanded}
\eeq
\end{widetext}
For the exchange term we make explicitly use of the normalization as $\vcrm{m}^2(y)=1$, $\frac{\rm d}{{\rm d}y}\,\vcrm{m}^2(y) = 2\,\vcrm{m}(y)\cdot\dot{\vcrm{m}}(y)=0$ and $ \frac{{\rm d}^2}{{\rm d}y^2}\,\vcrm{m}^2(y) = 2\:\vcrm{m}(y)\cdot\ddot{\vcrm{m}}(y) + 2\:\dot{\vcrm{m}}^2(y)=0$. 

Reminding that the distance $R_{ij}$ between atoms at site $i$ and $j$ is given by $(j-i)\Delta=R_{ij}$, replacing $\Delta$ by ${\rm d} y$ in (\ref{eq:spin-model_taylor-expanded}) in the limit of small changes, the energy functional of spin-model~(\ref{eq:spin-model}) approaches the energy functional of micromagnetic model~(\ref{eq:micmodEn}), $E\{\vcrm{S}\}\rightarrow E[\vcrm{m}]$, with parameters $A$, $\vcrm{D}$ and $\matx{K}$ as summarized in Eq.~(\ref{eq:ADK_connection}). 

\section{Spin-spiral solution of spin-lattice model with Dzyaloshinskii-Moriya interaction}
\label{app:sss-spin-lattice-model}

From the view-point of first-principles calculations of a magnetic crystalline solid, the planar helical or cycloidal spin-spiral represents an interesting magnetic state, because in the absence of spin-orbit coupling the spin-spiral state can be calculated by partitioning a solid into the same (chemical) unit cell that is used for non-magnetic or ferromagnetic calculations. 
This becomes possible by employing the generalized Bloch theorem~\cite{Sandratskii:91.1} and holds true for any arbitrary wave vector $\vcrm{q}\in\text{BZ}$ taken from the Brillouin zone (BZ) of wave vectors. 
It is a major concept to make such first-principles calculations feasible.

In this Appendix~\ref{app:sss-spin-lattice-model} we show that the planar homogeneous spin-spiral state of wave vector $\vcrm{q}$, whose rotation axis points parallel or antiparallel to the Dzyaloshinskii-Moriya vector, is a stationary solution, and for a particular wave vector $\vcrm{Q}$, the spin-spiral state is also the energy minimizer of the spin-lattice model~(\ref{eq:spin-model}) for a periodic solid, when the magnetic anisotropy term is ignored. 
It is known that the spin-spiral state is the stationary solution of a classical Heisenberg model on the Bravais lattice.\cite{Yoshimori:59, Villain:59, Kaplan:59, Kaplan:09} Here we show that the solution holds true also for the Heisenberg exchange plus DMI. 
In difference to the Heisenberg exchange only, where the energy is isotropic with respect to the rotation directions of the spirals, the DMI lowers the rotational symmetry, and selects spirals whose rotation directions are parallel and antiparallel to Dzyaloshinskii-Moriya vector $\vcrm{D}_{\vcrm{q}}$ of mode $\vcrm{q}$. 

In the following we assume a crystalline solid with lattice periodicity and restrict ourselves for simplicity to one atom per unit cell. 
We neglect the single-site anisotropy tensor in~(\ref{eq:spin-model}). 
The spin-model~(\ref{eq:spin-model}) on the Bravais lattice is then replaced by the quadratic form
\beq
  E\{\vcrm{S}\} = \frac{1}{2}\sum_{i,j}\vcrm{S}^\mathrm{T}_i \, \matx{J}_{ij}\, \vcrm{S}_j \komma
  \label{eq:spin-model_2}
\eeq
with prefactor 1/2 preventing a double counting of terms and the exchange tensor 
\beq
    \matx{J}_{ij}
  = \left( \begin{array}{ccc}
       \phantom{-} J_{ij}   & \phantom{-} D^z_{ij} &            -D^y_{ij} \\
                -  D^z_{ij} & \phantom{-} J_{ij}   & \phantom{-} D^x_{ij}\\
       \phantom{-} D^y_{ij} &            -D^x_{ij} & \phantom{-} J_{ij}
    \end{array} \right)\, \in \mathbb{R}^{3\times 3} \komma
  \label{eq:exchange-tensor}
\eeq
and $\matx{J}_{ij}=\matx{J}^\mathrm{T}_{ji}$. 
The lattice periodicity implies $\matx{J}_{ij}=\matx{J}_{0,j-i}=\matx{J}^\mathrm{T}_{0,i-j}$. 
The aim is to find the set of spins $\{\vcrm{S}_i\}$, with $\vcrm{S}_i: \mathbb{Z}\rightarrow\mathbb{S}^2\subset\mathbb{R}^3$, that minimizes $E\{\vcrm{S}\}$ subject to the constraints that the length of spins are on sphere $\mathbb{S}^2$ of radius $S$ and remain unchanged at all sites $i$,
\beq
  \vcrm{S}_i\cdot\vcrm{S}_i = S^2 \quad\forall\,i \in\mathbb{Z} \punkt
  \label{eq:constraint_strong}
\eeq
Luttinger and Tisza \cite{LuttingerTiszaPR1946,LuttingerPR1951} realized that the minimization of a quadratic form under $N$ strong constraints can be replaced by a much simpler problem of minimizing the energy~(\ref{eq:spin-model_2}) subject to the weak constraint
\beq
  \sum_i \vcrm{S}_i \cdot \vcrm{S}_i = N S^2 \komma
  \label{eq:constraint_weak}
\eeq
where $N$ is the number of lattice sites. 
This is a necessary condition and becomes sufficient if the solution also fulfills the strong constraint, as given in Eq.~(\ref{eq:constraint_strong}).

To take advantage of the translational symmetry of the crystalline solid, we transform 
the spin at lattice site $i$ with the lattice vector $\vcrm{R}_i$ into momentum space 
\beq
    \vcrm{S}_i
  = \frac{1}{\sqrt{N}}\sum_{\vcrm{q}}\vcrm{S}_{\vcrm{q}}\,e^{\imagi\vcrm{q}\vcrm{R}_i}
    \quad\text{and}\quad
    \vcrm{S}_{\vcrm{q}}
  = \frac{1}{\sqrt{N}}\sum_i\vcrm{S}_i\,e^{-\imagi\vcrm{q}\vcrm{R}_i} \punkt
\eeq 
Without loss of generality we assume here $\vcrm{R}_i\in\mathbb{R}^3$, but the derivations hold correct also for one- and two-dimensional lattices. 
Since $\vcrm{S}_i\in\mathbb{R}^3$ is a three-tuple of real numbers, it holds that $\vcrm{S}^\ast_{\vcrm{q}}=\vcrm{S}_{-\vcrm{q}}$. 
With this definition, the quadratic form~(\ref{eq:spin-model_2}) and the weak constraint~(\ref{eq:constraint_weak}) can be expressed in momentum space as 
\beq
    E\{\vcrm{S}_{\vcrm{q}}\}
  = \frac{1}{2}\sum_{\vcrm{q}} \vcrm{S}_{\vcrm{-q}}^\mathrm{T} \, \matx{J}_{\vcrm{q}} \, \vcrm{S}_{\vcrm{q}}
  = \frac{1}{2}\sum_{\vcrm{q}} \vcrm{S}_{\vcrm{q}}^\dagger     \, \matx{J}_{\vcrm{q}} \, \vcrm{S}_{\vcrm{q}}
  \label{eq:q-spin-model}
\eeq
and
\beq
    \sum_{\vcrm{q}} \vcrm{S}_{-\vcrm{q}} \cdot \vcrm{S}_{\vcrm{q}}
  = \sum_{\vcrm{q}} \vcrm{S}_{\vcrm{q}}^\dagger \,\vcrm{S}_{\vcrm{q}}
  = N S^2 \komma
  \label{eq:q-constraint-weak}
\eeq
respectively, with 
\bea
      \matx{J}^{\alpha\alpha^\prime}_{\vcrm{q}}
  &=& \sum_{j} \matx{J}^{\alpha\alpha^\prime}_{0j}\,\mrm{e}^{-\imagi\vcrm{q}(\vcrm{0}-\vcrm{R}_j)}
   =  \left(\matx{J}^{\alpha^\prime\alpha}_{\vcrm{q}}\right)^\ast 
   =  \left(\matx{J}^{\alpha\alpha^\prime}_{-\vcrm{q}}\right)^\ast  \nonumber\\
  &=& \begin{cases}
        \phantom{-} \left(\matx{J}^{\alpha\alpha^\prime}_{\vcrm{q}}\right)^\ast \quad\text{for}\> \alpha=  \alpha^\prime \\
                 -  \left(\matx{J}^{\alpha\alpha^\prime}_{\vcrm{q}}\right)^\ast \quad\text{for}\> \alpha\ne\alpha^\prime \\
      \end{cases} \komma 
  \label{eq:exchange-tensor-q-def}
\eea
and $\alpha \in \{ x,y,z\}$. 
Since the off-diagonal elements of $\matx{J}^{\alpha\alpha^\prime}_{\vcrm{q}}$ are purely imaginary, we replace $D^\alpha_{\vcrm{q}}$ by $\imagi D^\alpha_{\vcrm{q}}$.
The exchange tensor in momentum space is then related to the tensor in real space (\ref{eq:exchange-tensor}) as
\beq
    \matx{J}_{\vcrm{q}}
  = \left( \begin{array}{ccc}
      \phantom{-i}            J_{\vcrm{q}} & \phantom{-} \imagi\, D^z_{\vcrm{q}} &          -  \imagi\, D^y_{\vcrm{q}} \\
               -   \imagi\, D^z_{\vcrm{q}} & \phantom{-i}           J_{\vcrm{q}} & \phantom{-} \imagi\, D^x_{\vcrm{q}} \\
      \phantom{-}  \imagi\, D^y_{\vcrm{q}} &          -  \imagi\, D^x_{\vcrm{q}} & \phantom{-i}           J_{\vcrm{q}}
    \end{array} \right)\, \in \mathbb{C}^{3\times 3}  \komma
  \label{eq:exchange-tensor_momentum}
\eeq
with $J_{\vcrm{q}}$ and $D^\alpha_{\vcrm{q}}$ $\in \mathbb{R}$. 
Analogously, we find for the expression of the DM energy of the spin-model~(\ref{eq:spin-model})
\beq
  E_\mrm{DM}\{\vcrm{S}\} = \frac{1}{2} N \sum_{j} \vcrm{D}_{0j} \cdot \vcrm{C}_{0j}
  \quad\text{with}\quad
  \vcrm{C}_{0j}=\vcrm{S}_{0} \times \vcrm{S}_{j} \komma
  \label{eq:App_B_def_chirality}
\eeq
in terms of the vector chirality $\vcrm{C}_{0j}$, or sum over modes in momentum space, respectively,
\beq
    E_\mrm{DM}\{\vcrm{S}_{\vcrm{q}}\} = \frac{1}{2} \sum_{\vcrm{q}} \vcrm{D}_{\vcrm{q}}\cdot \vcrm{C}_{\vcrm{q}}
    \quad\text{with}\quad
    \vcrm{C}(\vcrm{S}_{\vcrm{q}}) = \imagi\vcrm{S}^\ast_{\vcrm{q}} \times \vcrm{S}_{\vcrm{q}} \komma
\eeq
where $\vcrm{C}(\vcrm{S}_{\vcrm{q}})$ is the vector chirality of mode $\vcrm{S}_{\vcrm{q}}$. 
Obviously energy is gained if the vector chirality is antiparallel to the Dzyaloshinskii-Moriya vector, $\vcrm{C}_{\vcrm{q}}\propto - \vcrm{D}_{\vcrm{q}}$.

To simplify the minimization problem it is convenient to transform the $3N$ dimensional quadratic form (\ref{eq:q-spin-model}) into the principal axes by diagonalizing the matrix $ \matx{J}_{\vcrm{q}}$. 
Since $\matx{J}_{\vcrm{q}} = \matx{J}_{\vcrm{q}}^\dagger$ is Hermitian, $ \matx{J}_{\vcrm{q}}$ has $3$ real eigenvalues $\lambda_{\vcrm{q},\gamma}$ with $\gamma\in\{1,2 ,3\}$: 
\beq
  \lambda_{\vcrm{q},1(3)} = J_{\vcrm{q}} \mP \abs{\vcrm{D}_{\vcrm{q}}}
  \quad\text{and}\quad
  \lambda_{\vcrm{q},2} = J_{\vcrm{q}}
  \label{eq:App_B_eigenvalues}
\eeq
with orthonormal eigenvectors $\matx{V}_{\vcrm{q}}=\{\vcrm{v}_{\vcrm{q},1}, \vcrm{v}_{\vcrm{q},2}, \vcrm{v}_{\vcrm{q},3}\}\in\mathbb{C}^{3\times 3}$. 
The eigenvector $\vcrm{v}_{\vcrm{q},2}$ points for each wave vector $\vcrm{q}$ into the direction of the Dzyaloshinskii vector $\vcrm{v}_{\vcrm{q},2}=\hat{\vcrm{e}}_{\vcrm{q},\mrm{DM}} =\vcrm{D}_{\vcrm{q}}/\abs{\vcrm{D}_{\vcrm{q}}}$ (see also Eq.~(\ref{eq:theta_D})). 
Obviously eigenvectors $\vcrm{v}_{\vcrm{q},1}$ and $\vcrm{v}_{\vcrm{q},3}$ live in the orthogonal subspaces. 
Without loss of generality we choose for each mode $\vcrm{q}$ the coordinate system of the spin space such that $\hat{\vcrm{e}}_{\vcrm{q},\mrm{DM}}$ coincides with the $z$-axis, $\hat{\vcrm{e}}_{\vcrm{q},\mrm{DM}} = \, \hat{\vcrm{e}}_{z}$. 
In this new frame of reference $\matx{R}_{\vcrm{q}}\vcrm{D}_{\vcrm{q}}=(0,0, D_{\vcrm{q}})^\mathrm{T}$ with $D_{\vcrm{q}} = \abs{\vcrm{D}_{\vcrm{q}}} \geq 0$, the eigenvectors transform to $\matx{W}_{\vcrm{q}}=\matx{R}_{\vcrm{q}}\matx{V}_{\vcrm{q}}=\{\vcrm{w}_{\vcrm{q},1}, \hat{\vcrm{e}}_z, \vcrm{w}_{\vcrm{q},3}\}\,\,\forall \vcrm{q}$, where $\matx{R}_{\vcrm{q}}\in\mathbb{R}^{3\times 3}$ is the respective rotation matrix, which conserves handedness, \ie, $\det{\matx{R}}=1$. 
With those definitions it is clear that $\matx{R}_{-\vcrm{q}} = -\matx{R}_{\vcrm{q}}$, because of the symmetry $\vcrm{D}_{-\vcrm{q}} = -\vcrm{D}_{\vcrm{q}}$ (see Eq.~(\ref{eq:exchange-tensor-q-def})) and our definition that the local $z$-axis points always parallel (not anti-parallel) to $\vcrm{D}_{\vcrm{q}}$.
We can always choose a transformation such that the exchange tensor becomes block diagonal
\beq
    \matx{R}_{\vcrm{q}}\,\matx{J}_{\vcrm{q}}\,\matx{R}_{\vcrm{q}}^\mathrm{T}
  = \left( \begin{array}{ccc}
                         J_{\vcrm{q}} &  \imagi D_{\vcrm{q}} & 0            \\
                 -\imagi D_{\vcrm{q}} &         J_{\vcrm{q}} & 0            \\
                         0            &         0            & J_{\vcrm{q}}  
    \end{array} \right) \komma 
  \label{eq:exchange-tensor-q2}
\eeq
the eigenvectors simplify to 
\beq
    \vcrm{w}_{\vcrm{q},1}
  = \frac{1}{\sqrt{2}} \left( \begin{array}{c} 1 \\ \imagi \\ 0 \end{array} \right)
  = \left( \vcrm{w}_{\vcrm{q},3} \right)^\ast \komma \quad  
    \vcrm{w}_{\vcrm{q},2}
  = \left( \begin{array}{c}  0 \\ 0 \\ 1 \end{array} \right) \komma
  \label{eq:Jq-eigenvects}
\eeq
and the energy in momentum space reads
\beq
    E\{\vcrm{S}_{\vcrm{q}}\}
  = \frac{1}{2}\sum_{\vcrm{q}}\sum_{\gamma=1}^3 
    \vcrm{S}_{\vcrm{q}}^\dagger \matx{R}^{\mrm{T}}_{\vcrm{q}} \vcrm{w}_{\vcrm{q},\gamma} 
    \lambda_{\vcrm{q},\gamma} \vcrm{w}^\dagger_{\vcrm{q},\gamma} \matx{R}_{\vcrm{q}} \vcrm{S}_{\vcrm{q}} \punkt
  \label{eq:q2-spin-model}
\eeq
The eigenvalues and eigenvectors are invariant with respect to space inversion symmetry $\matx{I}$ transforming $\matx{I}\vcrm{q}=-\vcrm{q}$, and complex valued functions as $\matx{I}\matx{J}(\vcrm{q}) = \matx{J}^\ast(-\vcrm{q})$, and $\matx{I}\vcrm{w}_{\vcrm{q},\gamma} = \vcrm{w}^\ast_{-\vcrm{q},\gamma}$. 
The matrix of vector chirality for the three eigenvectors $\matx{C}(\matx{W}_{\vcrm{q}})=\{-\hat{\vcrm{e}}_z,\vcrm{0}, \hat{\vcrm{e}}_z\}\,\,\forall \vcrm{q}$ are momentum independent and the chirality vector of the state with the lowest (highest) eigenvalue point antiparallel (parallel) to the direction of the DMI. 

Now we turn to our primary goal to find the state $\vcrm{S}_{\vcrm{Q}}$ that minimizes the energy expression $E\{\vcrm{S}_{\vcrm{q}}\}$ of Eqs.~(\ref{eq:q-spin-model}) or (\ref{eq:q2-spin-model}), respectively. 
Irrespective of the sign of $J_{\vcrm{q}}$, eigenvalue $\lambda_{\vcrm{q},1}$ is always the lowest eigenvalue for any wave vector $\vcrm{q}$,
$\Lambda_{\vcrm{q},\mrm{min}}= \lambda_{\vcrm{q},1}$. 
Considering the symmetry relation $\matx{J}(\vcrm{q}) = \matx{J}(-\vcrm{q})^\ast$ (see Eq.~\ref{eq:exchange-tensor-q-def})), both matrices have the same eigenvalues and subsequently $\Lambda_{\vcrm{q},\mrm{min}}=\Lambda_{-\vcrm{q},\mrm{min}}$ has at least a twofold degeneracy for $\vcrm{q}\in\text{BZ}$, but the eigenvectors corresponding to the lowest and highest value exchange their roles, \ie, $\vcrm{w}_{-\vcrm{q},1} = (\vcrm{w}_{\vcrm{q},1})^\ast = \vcrm{w}_{\vcrm{q},3}$ and \viceversa. 
A lower bound to $E\{\vcrm{S}_{\vcrm{q}}\}$ can be estimated considering that $\vcrm{S}_{\vcrm{q}}^\dagger \matx{J}_{\vcrm{q}} \vcrm{S}_{\vcrm{q}} \geq \vcrm{S}_{\vcrm{q}}^\dagger \Lambda_{\vcrm{q},\mrm{min}} \vcrm{S}_{\vcrm{q}}$ is limited by the lowest eigenvalue, and thus 
\beq
    E\{\vcrm{S}_{\vcrm{q}}\} \geq \frac{1}{2} \Lambda \sum_{\vcrm{q}} \vcrm{S}_{\vcrm{q}}^\dagger \, \vcrm{S}_{\vcrm{q}}
  = \frac{1}{2} \Lambda N S^2 \komma
  \label{eq:lower_bound}
\eeq
with $\Lambda = \Lambda_{\pm\vcrm{Q}} = \mathrm{min}_{\vcrm{q}} \Lambda_{\vcrm{q}} $ the lowest eigenvalue of all $\vcrm{q}$. 

In the state that minimizes the $3N$-dimensional ellipsoid transformed to the principal axis (see Eq.~(\ref{eq:q2-spin-model})) with respect to $\vcrm{S}^\dagger_{\vcrm{q}}$ subject to the constraint~(\ref{eq:q-constraint-weak}) of a $3N$-dimensional sphere of radius $NS^2$, it is easy to show by the method of Lagrange multipliers that the $\vcrm{S}_{\vcrm{q}}$ must satisfy 
\beq
    (\lambda_{\vcrm{q},\gamma} - 2\xi )\, \vcrm{w}^\dagger_{\vcrm{q},\gamma} \matx{R}_{\vcrm{q}} \vcrm{S}_{\vcrm{q}}
  = 0 \quad \forall \vcrm{q}, \gamma \komma
  \label{eq:lagrangeparameter}
\eeq
where $\xi$ is the Lagrange multiplier independent of $\vcrm{q}$. 
Using Eqs.~(\ref{eq:lagrangeparameter}) and (\ref{eq:q-constraint-weak}), the energy~(\ref{eq:q2-spin-model}) becomes
\beq
  E\{\vcrm{S}_{\vcrm{q}}\} = \xi N S^2 \punkt
\eeq
Hence the minimum $E\{\vcrm{S}_{\vcrm{Q}}\}$ is obtained for the minimum $\xi$ for which solutions of Eq.~(\ref{eq:lagrangeparameter}) exist, which is realized for $\xi =1/2\Lambda$, proving that the ground state satisfies the equal sign in Eq.~(\ref{eq:lower_bound}).

Now we have a closer look at Eq.~(\ref{eq:lagrangeparameter}). 
If $\vcrm{S}_{\vcrm{q}} \ne {\vcrm{0}}$ for all $\vcrm{q}$ and since $\matx{W}_{\vcrm{q}}$ spans the whole three-dimensional spin-space there exists for each vector $\vcrm{q}$ at least one eigenvector $\vcrm{w}_{\vcrm{q},\gamma}$ for which $\vcrm{w}^\dagger_{\vcrm{q},\gamma} \matx{R}_{\vcrm{q}} \vcrm{S}_{\vcrm{q}} \ne {\vcrm{0}}$. 
As a consequence for each vector $\vcrm{q}$, there is at least one eigenvalue $\lambda_{\vcrm{q},\gamma}$ for which $\lambda_{\vcrm{q},\gamma}=2\xi$. 
Since $\xi$ is independent of $\vcrm{q}$, all eigenvalues and thus all $J_{\vcrm{q}}$, and $\abs{\vcrm{D}_{\vcrm{q}}}$ should be independent of $\vcrm{q}$, and this is unphysical. 
On the contrary a single-q state, 
\beq
    \vcrm{S}_{\vcrm{q} \vert \bar{\vcrm{q}},\gamma}
  = \sqrt{\frac{N S^2}{2}} ~ \matx{R}^{\mrm{T}}_{\bar{\vcrm{q}}} \,
    \left(   \vcrm{w}_{\bar{\vcrm{q}},\gamma}  \delta_{\vcrm{q},\bar{\vcrm{q}}}
           + \vcrm{w}_{-\bar{\vcrm{q}},\gamma} \delta_{\vcrm{q},-\bar{\vcrm{q}}} \right)\komma
  \label{eq:single-q_state}
\eeq
\ie, a state for which $\vcrm{S}_{\vcrm{q}} = {\vcrm{0}},\, \forall\,\vcrm{q}\setminus\{\bar{\vcrm{q}},-\bar{\vcrm{q}}\}$, made of a superposition of two arbitrary modes with wave vectors $\bar{\vcrm{q}}$ and $-\bar{\vcrm{q}}$, for which eigenvalue $ \lambda_{\bar{\vcrm{q}},\gamma}=\lambda_{-\bar{\vcrm{q}},\gamma}$ is two-fold degenerate, with polarization directions determined by the principal axes of the exchange tensor $\matx{J}_{\bar{\vcrm{q}}}$ satisfies Eq.~(\ref{eq:lagrangeparameter}) for the Lagrange parameter $\xi=1/2 \lambda_{\bar{\vcrm{q}},\gamma}$ and the respective energy 
\beq
    E\{\vcrm{S}_{\vcrm{q} \vert \bar{\vcrm{q}},\gamma}\}
  = \frac{1}{2}\lambda_{\bar{\vcrm{q}},\gamma}NS^2 \komma
  \label{eq:single-q_state-energy}
\eeq 
and it is a stationary state of the energy functional, with $\lambda_{\bar{\vcrm{q}},\gamma}$ from Eq.~(\ref{eq:App_B_eigenvalues}). 
The term $\propto \matx{R}^{\mrm{T}}_{\bar{\vcrm{q}}} \vcrm{w}_{\bar{\vcrm{q}},\gamma} \delta_{\vcrm{q},\bar{\vcrm{q}}} $ satisfies also condition~(\ref{eq:lagrangeparameter}), but not the condition $\vcrm{S}^\ast_{\vcrm{q}}=\vcrm{S}_{-\vcrm{q}}$, and thus $\vcrm{S}_i\notin\mathbb{R}$. 
Therefore, this case is not further discussed.

The three eigenmodes~(\ref{eq:single-q_state}) exhibit the chirality
\bea
      \vcrm{C}(\vcrm{S}_{\vcrm{q} \vert \bar{\vcrm{q}},1})
  &=& -\frac{N S^2}{2} \, \matx{R}^{\mrm{T}}_{\bar{\vcrm{q}}} \hat{\vcrm{e}}_z
      \left( \delta_{\vcrm{q},\bar{\vcrm{q}}} - \delta_{\vcrm{q},-\bar{\vcrm{q}}}\right) \\
      \vcrm{C}(\vcrm{S}_{\vcrm{q} \vert \bar{\vcrm{q}},2})
  &=& \phantom{-} \vcrm{0} \\
      \vcrm{C}(\vcrm{S}_{\vcrm{q} \vert \bar{\vcrm{q}},3})
  &=& \phantom{-} \frac{N S^2}{2} \, \matx{R}^{\mrm{T}}_{\bar{\vcrm{q}}} \hat{\vcrm{e}}_z
      \left( \delta_{\vcrm{q},\bar{\vcrm{q}}} - \delta_{\vcrm{q},-\bar{\vcrm{q}}}\right) \punkt
\eea
Obviously, the modes $\vcrm{S}_{\vcrm{q} \vert \bar{\vcrm{q}},1}$ and $\vcrm{S}_{\vcrm{q} \vert \bar{\vcrm{q}},3}$ are of opposite chirality.

A Fourier back-transformation of the eigenmodes shows that the modes with $\gamma=1$ and $\gamma=3$ correspond to flat spin spirals,
\beq
    \vcrm{S}_{i \vert \bar{\vcrm{q}},1(3)}
  = S \, \matx{R}^{\mrm{T}}_{\bar{\vcrm{q}}}
    \left( \begin{array}{c} \phantom{-} \cos(\bar{\vcrm{q}} \cdot \vcrm{R}_i) \\
                            \mP         \sin(\bar{\vcrm{q}} \cdot \vcrm{R}_i) \\
                            0
           \end{array} \right) ~ \komma
\eeq
where the upper (lower) sign corresponds to mode 1 (3) and the rotation corresponds to a clockwise (counter-clockwise) rotation around $\hat{\vcrm{e}}_{\bar{\vcrm{q}},\mrm{DM}} = \matx{R}^{\mrm{T}}_{\bar{\vcrm{q}}} \hat{\vcrm{e}}_z$. 
The assignment of sign and handedness is consistent (i) with the common definition of the rotation matrix
\beq
    \matx{R}_z(\varphi)
  = \left(\begin{array}{ccc}
    \phantom{\pM} \cos\varphi & \pM          \sin\varphi & 0 \\
    \mP           \sin\varphi & \phantom{\pM}\cos\varphi & 0 \\
    0                         & 0                        & 1 \\
    \end{array} \right)
\eeq
rotating vector $\vcrm{S}_{i \vert \bar{\vcrm{q}},1(3)}$ in a right-handed coordinate system clockwise (counter-clockwise) around $\hat{\vcrm{e}}_z$ by an angle $\varphi=\bar{\vcrm{q}} \cdot \vcrm{R}_i$ with $0\le\varphi\le \pi$, (ii) as well as with the definition of the winding number
\bea
      w_{1(3)}
  &=& \frac{1}{2\pi} \oint \intover{}{\varphi}
      \,\,\vcrm{S}_{1(3)}\times\tdd{\vcrm{S}_{1(3)}}{\varphi} \nonumber\\
  &=& \frac{1}{2\pi}\oint \intover{}{\varphi}
                   \left[  S_{x\vert 1(3)}(\varphi)\tdd{S_{y\vert 1(3)}(\varphi)}{\varphi} \right. \nonumber\\
  & & \hspace{4em} \left. -S_{y\vert 1(3)}(\varphi)\tdd{S_{x\vert 1(3)}(\varphi)}{\varphi} \right] \nonumber\\
  &=&              \mP 1
\eea
counting the total number of turns of the spin-spiral as a curve parametrized by $\varphi$ with $0\le\varphi\le 2\pi$, where counter-clockwise motion counts as positive and clockwise motion counts as negative integers, and (iii) with the vector spin-chirality between atom $i$ and $i+1$ defined in
\beq
    \vcrm{C}_{i,i+1}(\vcrm{S}_{i \vert \bar{\vcrm{q}},1(3)})
  = \mP S^2 \sin\left( \bar{\vcrm{q}} \left( \vcrm{R}_{i+1} - \vcrm{R}_{i} \right) \right) \,
    \hat{\vcrm{e}}_{\bar{\vcrm{q}},\mrm{DM}} \punkt
\eeq
Alternatively we could also say, that mode 1 (3) rotates counter-clockwise (clockwise) around $\mP \hat{\vcrm{e}}_{\bar{\vcrm{q}},\mrm{DM}}$, but this is not the definition we follow here. 
These two modes $\gamma=1,3$ are separated by an energy $N S^2 \abs{\vcrm{D}_{\bar{\vcrm{q}}}}$.

On the contrary, the mode $\vcrm{S}_{\vcrm{q} \vert \bar{\vcrm{q}},2}$ represents a spin density wave in the direction of $\vcrm{D}_{\bar{\vcrm{q}}}$,
\beq
    \vcrm{S}_{i \vert \bar{\vcrm{q}},2}
  = S \,\hat{\vcrm{e}}_{\bar{\vcrm{q}},\mrm{DM}} \,\cos(\bar{\vcrm{q}} \cdot \vcrm{R}_i)~ \punkt
\eeq
All three modes satisfy per construction the weak constraint~(\ref{eq:constraint_weak}), but this mode does not fulfill the strong constraint~(\ref{eq:constraint_strong}) and thus must be excluded from the set of solutions. 

$\bar{\vcrm{q}}$ takes the physical meaning of the propagation vector of the spin spiral. 
If the propagation vector is parallel to the DMI-vector, $\bar{\vcrm{q}}\Vert\vcrm{D}_{\bar{\vcrm{q}}}$, then we call the spiral a helical or Bloch-type spin-spiral for which holds that $\curl{\vcrm{S}_{\bar{\vcrm{q}},1(3)}}=\nabla_{\vcrm{R}}\times \vcrm{S}_{\bar{\vcrm{q}}1(3)}=\pM(\bar{\vcrm{q}}\cdot\hat{\vcrm{e}}_{\bar{\vcrm{q}},\mrm{DM}}) \vcrm{S}_{\bar{\vcrm{q}}1(3)}$. 
If the propagation vector is perpendicular to the DMI-vector, $\bar{\vcrm{q}}\perp\vcrm{D}_{\bar{\vcrm{q}}}$, then we name the spiral a cycloidal or N\'eel-type spin-spiral for which $\curl{\vcrm{S}_{\bar{\vcrm{q}},1(3)}}= \mP(\bar{\vcrm{q}}\cdot\vcrm{S}_{\bar{\vcrm{q}},1(3)})\hat{\vcrm{e}}_{\bar{\vcrm{q}},\mrm{DM}}$. 
Here the spin-spiral $\vcrm{S}_{\bar{\vcrm{q}}}=\vcrm{S}_{\bar{\vcrm{q}}}(\vcrm{R}):\mathbb{R}^3\rightarrow\mathbb{S}^2$ is a smooth function whose values coincide at the positions of the lattice vectors $\vcrm{R}_i$ with $\vcrm{S}_{i \vert \bar{\vcrm{q}}}$. 
The details of the vector relation between $\bar{\vcrm{q}}$ and $\vcrm{D}_{\bar{\vcrm{q}}}$ depend on the symmetry of the crystal lattice.

So far we focused on physical realizations where the eigenvalues of the stationary state $\Lambda_{\bar{\vcrm{q}}}$ are exactly two-fold degenerate, namely for $\bar{\vcrm{q}}$ and $-\bar{\vcrm{q}}$ for $\bar{\vcrm{q}}, -\bar{\vcrm{q}} \in\text{BZ}$. 
For systems with a non-trivial point group, the $\bar{\vcrm{q}}$-vector is equivalent to a star of $p$ $\bar{\vcrm{q}}$-vectors, $\{\bar{\vcrm{q}}\}$, formed by consideration of $\bar{\vcrm{q}}_\alpha=\matx{P}_{\alpha}\bar{\vcrm{q}}$ for all symmetry operations $\alpha$ denoted by $\matx{P}_\alpha$ of the symmetry group of the lattice. 
Accordingly, $\Lambda_{\{\bar{\vcrm{q}}\}}$ is $p$-fold degenerate, $p$ different $\vcrm{S}_{\bar{\vcrm{q}}_\alpha}$ can satisfy Eq.~(\ref{eq:lagrangeparameter}) simultaneously, and the single-q state $\vcrm{S}_{\vcrm{q} \vert \bar{\vcrm{q}},\gamma}$ in Eq.~(\ref{eq:single-q_state}) may be replaced by an alternative ansatz describing a multi-q state by a superposition of properly normalized $\vcrm{S}_{\vcrm{q} \vert \bar{\vcrm{q}}_\alpha,\gamma}$ for any choice of $\alpha$ taken from the symmetry group as long as the strong constraint~(\ref{eq:constraint_strong}) is fulfilled. 
For $\bar{\vcrm{q}}={\vcrm{Q}}$ we may expect a multi-q ground state. 
The competition of the various possible multi-q states with the single-q state as possible ground state is typically determined by energy contributions beyond the model discussed here.\cite{Kurz:01}

\section{Extracting micromagnetic parameters from first-principles energetics of spin-spiral state}
\label{app:sss-connection-parameter}

An important aspect in undertaking multi-scale simulations of magnetic structures is the development of realistic micromagnetic models with material-specific parameters. 
Here, we show that the spin stiffness $A$ and the spiralization $\vcrm{D}$, which enter the micromagnetic model (see Eq.~(\ref{eq:micmodEn})) can be extracted directly from first-principles calculations of the total energy $E_\mrm{tot}(\vcrm{q},\hat{\vcrm{e}}_\mrm{rot})$ given per magnetic atom for a planar homogeneous spin-spiral with wave vector $\vcrm{q}$ and fixed rotation axis $\hat{\vcrm{e}}_\mrm{rot}$ related to the planar spiral as $\hat{\vcrm{e}}_\mrm{rot}=\hat{\vcrm{m}}_i\times \mrm{d}\hat{\vcrm{m}}_i/\mrm{d}\vcrm{R}_i$.

In the following we give a rather general derivation not restricted to one-dimensional chains and thus we work with the spin-stiffness tensor $\matx{A}\in \mathbb{R}^{d\times d}$ rather than with the spin stiffness $A$ and the spiralization tensor $\matx{D}\in \mathbb{R}^{3\times d}$, also called matrix of Dzyaloshinskii-Moriya constants, rather than the Dzyaloshinskii-vector $\vcrm{D}$. 
$d\in \{1,2,3\}$ refers to the dimensionality of the micromagnetic problem with $d=1$ relevant for chains, domain-walls or magnetic spirals, $d=2$ for films, interfaces or the treatment of skyrmions and $d=3$ for bulk or bubbles for example. 

In the general case, the expression of the DM energy density of the micromagnetic energy functional~(\ref{eq:micmodEn}) translates from the one-dimensional case $\Dvec\cdot\left(\vcrm{m}\times\dot{\vcrm{m}}\right)$ to $\matx{D}\! :\!\left(\vcrm{m}\times\nabla\vcrm{m}\right)$ in case of higher dimensions. 
The expression in the parenthesis is also called the Lifshitz matrix $\matx{L}(\vcrm{m})\in \mathbb{R}^{3\times d}\subseteq\mathbb{R}^{3\times 3}$, a matrix of differential one-forms with matrix elements $L_{\alpha\beta}= \sum_{\alpha^\prime\alpha^{\prime\prime}} \epsilon_{\alpha\alpha^\prime\alpha^{\prime\prime}} L_{\alpha^\prime\alpha^{\prime\prime}}^{(\beta)}$, with $\alpha, \beta\in \{x, y, z\}$. 
$L_{\alpha\alpha^\prime}^{(\beta)}=-( m_\alpha\partial_\beta m_{\alpha^\prime}-m_{\alpha^\prime}\partial_\beta m_\alpha)$ are the Lifshitz invariants and $\epsilon$ is the Levi-Civita symbol. 
The operator ``:'' refers to the inner product of two matrices: 
$\matx{D}\! :\! \matx{L}(\vcrm{m}) = \trace{\matx{D}^{\rm T}\matx{L}}=\sum_{\alpha\beta} D_{\alpha\beta}L_{\alpha\beta}$. 
Which of the Lifshitz invariants or which of the maximal 9 components of the $\matx{D}$-matrix, respectively, are relevant depends on the point-group of the crystal, an aspect which is not considered here any further.

Starting point of the derivation is the observation made in Appendix~\ref{app:sss-spin-lattice-model} that flat spirals with rotation axis $\mp\hat{\vcrm{e}}_\mrm{DM}(\vcrm{q})$ are the stationary states of the spin-lattice model with Heisenberg and DM interaction with an energy per atom of $E(\vcrm{q},\mP\hat{\vcrm{e}}_\mrm{DM}(\vcrm{q}))=\frac{1}{2}\lambda_{\vcrm{q},1(3)} = \frac{1}{2}(J(\vcrm{q}) \mP \abs{\vcrm{D}(\vcrm{q})})$ and two opposite rotation senses.\footnote{That $\lambda_{\vcrm{q},\gamma}$ corresponds to an energy per magnetic atom can be inferred from Eq.~(\ref{eq:single-q_state-energy}).} 
The extraordinary nature of these states lies in the efficient realization in first-principles theory of the electronic structure. 
We recall from the discussion in Sec.~\ref{sec:chMethod} that by neglecting the magnetic anisotropy energy, the total energy of a spin-spiral state, $E_\mrm{tot}(\vcrm{q},\hat{\vcrm{e}}_\mrm{rot})\simeq E_\mrm{SS}(\vcrm{q})+ E_\mrm{DM}(\vcrm{q},\hat{\vcrm{e}}_\mrm{rot})$, can be explicitly approximated by calculations in two steps: 
self-consistent calculations of the spin-spiral energy without SOC, $E_\mrm{SS}(\vcrm{q})$, and the energy due to SOC, $E_\mrm{DM}(\vcrm{q},\hat{\vcrm{e}}_\mrm{rot})$, in first order perturbation theory for a given rotation axis $\hat{\vcrm{e}}_\mrm{rot}$. 
We work here with a normalized length of magnetic spins, $S^2=1$, as typical for spin-lattice and micromagnetic models. 
The parameters $J(\vcrm{q})$ and $\vcrm{D}(\vcrm{q})$ of particular systems are then obtained from first principles by a direct comparison of the energies $E(\vcrm{q},\mP\hat{\vcrm{e}}_\mrm{DM}(\vcrm{q}))=\frac{1}{2}\lambda_{\vcrm{q},1(3)}$,
\bea
  \label{eq:appC_Jq}
  \delta J(\vcrm{q})      =  J(\vcrm{q}) - J(\vcrm{0})
                         &=& 2 \left( E_\mrm{SS}(\vcrm{q}) - E_\mrm{SS}(\vcrm{0}) \right)\qquad \\
  \hspace{-0.5cm} \text{and} \qquad\qquad\qquad
  | \vcrm{D}(\vcrm{q}) | &=& 2 E_\mrm{DM} \left(\vcrm{q},\hat{\vcrm{e}}_\mrm{DM} \right) \komma
  \label{eq:appC_Dq}
\eea
$\forall \vcrm{q}\in\text{BZ}$. 
Note, the equalities hold only if $E_\mrm{SS}$ and $E_\mrm{DM}$ are given as energy per magnetic atom.
The back Fourier transformation of $J(\vcrm{q})$ and $\vcrm{D}(\vcrm{q}) $ according to Eq.~(\ref{eq:exchange-tensor-q-def}) provides then the Heisenberg exchange parameter $J_{0j}$ and the microscopic DM-vectors $\vcrm{D}_{0j}$ on the real space lattice. 
There might be cases where the direction $\hat{\vcrm{e}}_\mrm{DM}(\vcrm{q})$ is not known \apriori. 
In this case we can determine the three components of $\vcrm{D}(\vcrm{q})$ applying Eq.~(\ref{eq:appC_Dq}) for each wavevector $\vcrm{q}$ and for three independent axes of spiral rotations $\hat{\vcrm{e}}_\mrm{rot}$, \ie, $\hat{\vcrm{e}}_\mrm{rot}\cdot \vcrm{D}(\vcrm{q}) = 2 E_\mrm{DM} \left(\vcrm{q},\hat{\vcrm{e}}_\mrm{rot} \right)$.

In micromagnetic models one typically assumes that the ground state of the system is close to the collinear state $\vcrm{q}_\mrm{c}$, \eg, the ferromagnetic state at $\vcrm{q}_\mrm{c}=\vcrm{0}$ or an antiferromagnetic state at a high-symmetry point at the boundary of the Brillouin zone. 
At such a point $\vcrm{q}_\mrm{c}$ in the Brillouin zone, $\abs{J(\vcrm{q}_\mrm{c})}$ typically takes a local minimum. 
If we assume that $\abs{{D}(\vcrm{q})}\ll \abs{J(\vcrm{q})}$ for $\vcrm{q}$ in the vicinity of $\vcrm{q}_\mrm{c}$ and measured from $\vcrm{q}_\mrm{c}$, \ie, the long wavelength limit where $|\vcrm{q}|$ is small, the relevant energy landscape may be explored by Taylor expanding Eqs.~(\ref{eq:appC_Jq}) and (\ref{eq:appC_Dq}), \ie, the exchange parameters
\bea
  \hspace{-0.5cm}
                    \delta J(\vcrm{q})
  \!&\!=\!&\!       2 \sum_{j \geq 1}{ J_{0j} \, (\cos(\vcrm{q} \vcrm{R}_j) } -1) \nonumber\\
  \!&\!\approx\!&\! - \sum_{j \geq 1}{ J_{0j} ( \vcrm{R}_j\vcrm{q})^2 }\!\!\komma
  \label{eq:app_exch_parameter_J}\quad\\
  \hspace{-0.5cm}
                    D(\vcrm{q})\,\hat{\vcrm{e}}_\mrm{DM}(\vcrm{q}) 
  \!&\!=\!&\!       D(\vcrm{q})\,\sum_{k=1}^3 (\hat{\vcrm{e}}_\mrm{DM}(\vcrm{q})
                      \cdot \hat{\vcrm{e}}^{(k)}_\mrm{rot})\, \hat{\vcrm{e}}^{(k)}_\mrm{rot}\nonumber\\
  \!&\!=\!&\!       2 \sum_{j \geq 1}{ \vcrm{D}_{0j}  \sin(\vcrm{q} \vcrm{R}_j) } \nonumber\\
  \!&\!\approx\!&\! 2 \sum_{j \geq 1}{\vcrm{D}_{0j} \,( \vcrm{R}_j\vcrm{q})}\quad
  \label{eq:app_exch_parameter_D}
\eea
and the total energy
\beq
    \frac{1}{V_\Omega}E_\mrm{tot}(\vcrm{q},\hat{\vcrm{e}}_\mrm{rot})
  =   \left[\hat{\vcrm{e}}_\mrm{rot}\cdot\frac{\matx{D}}{2\pi}\right]^\mrm{T}\!\! \vcrm{q}
    + \vcrm{q}^\mrm{T}\frac{\matx{A}}{4\pi^2} \, \vcrm{q} +\mathcal{O}(q^3)
  \label{eq:app_totalenergyexpansion}
\eeq 
up to second order in $\abs{\vcrm{q}}$ measured relative to $\vcrm{q}_\mrm{c}$, for a fixed rotation axis $\hat{\vcrm{e}}_\mrm{rot}$. 
$k$ labels the maximally three linear independent rotation axes $\hat{\vcrm{e}}^{(k)}_\mrm{rot}$. 
In Eqs.~(\ref{eq:app_exch_parameter_J}) and (\ref{eq:app_exch_parameter_D}) we made explicitly use of the symmetry relations $J(\vcrm{q})=J(-\vcrm{q})$ and $\vcrm{D}(\vcrm{q}) = - \vcrm{D}(-\vcrm{q})$ and chose $J(\vcrm{q}_\mrm{c})$ as origin of energy.
In Eq.~\eqref{eq:app_totalenergyexpansion}, $V_\Omega$ represents the volume of the unit cell in a 3D system, the surface area in 2D, or, respectively, the spacing between magnetic atoms in 1D, so that the left-hand side represents an energy density. 
The numerical prefactors appear in order to keep consistency with the definition of the energy functional in the main text, Eq.~(\ref{eq:micmodEn}). 
We arrive at the energy expression with the spin stiffness $\matx{A}$ obtained from the curvature of the total energy at wave vector $\vcrm{q}_\mrm{c}$,
\bea
\frac{\matx{A}}{4\pi^2}
&=& \frac{1}{2V_\Omega} \frac{\partial^2 }{ \partial \vcrm{q}^2} E_\mrm{SS}(\vcrm{q})|_{\vcrm{q}_c} \nonumber\\
&=& \frac{1}{2V_\Omega} \sum_{j \geq 1} J_{0j}\vcrm{R}_j \otimes \vcrm{R}_j\punkt
\eea
The projection of the spiralization matrix $\matx{D}$ onto the direction of the rotation vector $\hat{\vcrm{e}}_\mrm{rot}$, $[\hat{\vcrm{e}}_\mrm{rot}\cdot\matx{D}]\in \mathbb{R}^{d}$, is obtained from the slope of the total energy calculated for the rotation axis $\hat{\vcrm{e}}_\mrm{rot}$,
\bea
      \frac{\matx{D}}{2\pi}
  &=& \hat{\vcrm{e}}_\mrm{rot}\otimes\frac{1}{V_\Omega}
      \frac{\partial}{\partial\vcrm{q}} E_\mrm{DM}(\vcrm{q},\hat{\vcrm{e}}_\mrm{rot})|_{\vcrm{q}_c} \nonumber\\
  &=& \frac{1}{V_\Omega} \sum_{j \geq 1}\vcrm{D}_{0j}\otimes \vcrm{R}_j\punkt\quad
  \label{eq:erot_dot_D}
\eea
The product denoted by "$\otimes$" indicates the tensorial property of the spin-stiffness and the DM-matrix.
These equations provide the link between the micromagnetic parameters, the \abinitio results of spin-spiral calculations and a spin-lattice model. 
On the basis of the first expression on the right-hand side of Eq.~(\ref{eq:erot_dot_D}) we can interpret the $\matx{D}$-matrix as a tensor in the space spanned by the magnetization direction and real space.

\section{Atom-dependent micromagnetic $\Dvec$-vector in the Fert-Levy model obtained for a spin-spiral state}
\label{app:sss-Fert-Levy-Model}

Applying the Fert-Levy model~\cite{FL80,LF81} for determining the microscopic DM vector in metals is very appealing due to its simplicity and clarity. 
It is frequently applied to determine the direction of the DM vector and interesting because it could provide a basis for interpreting our first-principles results. 
The assumptions under which the model is derived motivates this appendix, where we explore the validity and applicability of the model to extended periodic systems, primarily to chains.

Fert and Levy investigated CuMnT ternary alloys, with a small concentration (1\% or 2\%) of Mn impurities and of heavy non-magnetic atoms T$=$Au or Pt. 
They found that experimental anisotropy data are explained by Mn atoms carrying a magnetic moment and interacting not only by the typical Rudermann-Kittel-Kasuya-Yosida (RKKY) interaction, but also by a DM-type interaction due to spin-orbit scattering of the conduction electrons by the non-magnetic impurities. 
They derived the leading order expression for the DMI-energy that is first order in the spin-orbit coupling and second order in the exchange interaction, which results from an expression for the third-order perturbation of the ground-state energy of the gas of conduction electrons due to the presence of the Mn spins and the non-magnetic impurity. 
Evaluating this expression under the assumption that 
($i$)   the magnetic moments are located at the Mn atoms and 
($ii$)  the spin-orbit interaction at the non-magnetic impurity atoms only, 
($iii$) that both atom types are located in the host as impurities in the low, but not very low ($>1000$~ppm), concentration limit, 
($iv$)  that Cu provides the gas of homogeneous electrons described by the Fermi energy, $E_\mrm{F}$, and wavevector, $k_\mrm{F}$, that
($v$)   scatter at the non-magnetic impurity with the scattering phase shift $\delta_2(E_\mrm{F})$ and the spin-orbit strength $\lambda$, and hybridize with the $3d$ states of the Mn atoms described by $\Gamma$, the exchange interaction strength between the host electrons and local spins, they arrived at a trilinear expression for the DMI-energy,
\begin{widetext}
\beq
    E^{ij\mu}_\mrm{DM}
  = -\frac{135 \pi}{32} \frac{\lambda \,\Gamma^2}{E_\mrm{F}^2 \,k_\mrm{F}^3}
    \sin\delta_2(E_\mrm{F}) \ \sin[k_\mrm{F}(R_{i\mu}+R_{j\mu}+R_{ij})+\delta_2(E_\mrm{F})] \\ 
    \cdot \frac{\hat{\vcrm{R}}_{i\mu} \cdot \hat{\vcrm{R}}_{j\mu}}{R_{i\mu} R_{j\mu} R_{ij}}
    (\hat{\vcrm{R}}_{i\mu}\times\hat{\vcrm{R}}_{j\mu})
    \cdot (\vcrm{S}_{i} \times \vcrm{S}_{j}) \komma
\eeq
\end{widetext}
relating three atoms: 
one non-magnetic impurity denoted by $\mu$ and two magnetic impurities denoted by $i,j$ placed at the position $\vcrm{R}_{i(j)\mu}$ measured from the position of the spin-orbit impurity. 
$\vcrm{R}_{ij}$ measures the distance between the two atoms $i$ and $j$.

Now we apply this model to a single transition-metal chain on the Pt substrate. 
We are aware of the fact that neither the spin-orbit atoms Pt nor the magnetic chain atoms are in the low concentration limit. 
Also the Fermi surface of Pt is more complex than Cu and the isotropic approximation of the Fermi wavevector underlying this model is a further approximation. 
Further, we assume that all Pt atoms, irrespective of their distance and position from the chain are electronically identical, \ie, $\delta_2^\mu = \delta_2$, and $\lambda^\mu=\lambda$. 
In difference to the \abinitio calculations we consider here a truly single magnetic chain and no periodic repetition of the surface unit cell. 
This is a good approximation for Pt atoms $\mu$ close to the chains, but differences are expected for atoms in the center of the terrace as they experience competitive interactions to chains at the upper and lower terrace.

The DM-energy contribution of a magnetic texture which arises solely through the presence of a certain spin-orbit atom $\mu$ is given by
\beq
  E_\mrm{DM}^\mu = \sum_{\langle i,j\rangle} E^{ij\mu}_\mrm{DM} \punkt
  \label{eq:fertlevy_doublesum}
\eeq
The brackets denote a summation over all pairs of magnetic chain-atoms $i$ and $j$. 
For a fixed atom $\mu$, the direction of $\hat{\vcrm{R}}_{i\mu}\times\hat{\vcrm{R}}_{j\mu}$ in Eq.~(\ref{eq:fertlevy_doublesum}) is always the same, irrespective of $i$ and $j$, and we denote it by $\hat{\vcrm{n}}_{\mu} = \hat{\vcrm{e}}_y \times \hat{\vcrm{d}}_\mu$. 
Here, $\hat{\vcrm{d}}_\mu$ is the unit vector pointing from the atom $\mu$ into the direction of shortest distance to the chain. 
Furthermore, we define $\varphi_{i\mu}$ as the angle between $\hat{\vcrm{R}}_{i\mu}$ and $\hat{\vcrm{d}}_\mu$.

In the spirit of the first-principles calculations, we next consider a homogeneous spin spiral for which $\vcrm{S}_{i} \times \vcrm{S}_{j} = \sin(q \, a_y\,(j-i)) \, \hat{\vcrm{e}}_\mrm{rot}$, where $\hat{\vcrm{e}}_\mrm{rot}$ is the rotation axis, and we obtain
\begin{widetext}
\bea
      E^\mu_{\mrm{DM}}(q)
  &=& - C ~ \varepsilon_{\mu}(q) ~ \hat{\vcrm{n}}_{\mu} \cdot \hat{\vcrm{e}}_\mrm{rot}
      \qquad \text{with} \qquad
      C = \frac{135 \pi}{32} \frac{\lambda \,\Gamma^2}{E_\mrm{F}^2 \,k_\mrm{F}^3} \sin\delta_2(E_\mrm{F})
      \qquad \text{and} \\
      \varepsilon_{\mu}(q)
  &=& \sum_{\langle i,j \rangle} \sin[k_\mrm{F}(R_{i\mu}+R_{j\mu}+R_{ij})+\delta_2(E_\mrm{F})] 
      \cdot\frac{\cos(\varphi_{j\mu} - \varphi_{i\mu})}{R_{i\mu} R_{j\mu} R_{ij}} 
      ~\sin(\varphi_{j\mu} - \varphi_{i\mu}) ~ \sin(q \, a_y\,(j-i)) \punkt
  \label{eq:FL:epsilon_doublesum}
\eea
\end{widetext}
Clearly, $E^\mu_\mrm{DM}$ is lowest if $\hat{\vcrm{e}}_\mrm{rot}$ is parallel (anti-parallel) to $\hat{\vcrm{n}}_{\mu}$, depending on the sign of the prefactor.

The contribution of this atom to the spiralization is defined as
\beq
    \frac{\vcrm{D}_\mu}{2\pi}
  = \left. \frac{1}{a_y} \frac{\partial E^\mu_\mrm{DM}}{\partial q} \right\rvert_{q=0}
    \cdot \hat{\vcrm{n}}_{\mu} \punkt
  \label{eq:FL:Eq_derivative}
\eeq
Unfortunately, $\vcrm{D}_\mu$ diverges in the limit $q \rightarrow 0 $ for this periodic model for one dimension, due to an effective $1/R$-dependence for each of the two sums over $i$ and $j$ contained in Eq.~(\ref{eq:FL:Eq_derivative}). 
In a realistic solid we expect a finite phase coherence length of the wave function or some structural or chemical disorder, which truncates the interaction range of the atoms in the finite chain and thus the summations in the sums in Eq.~(\ref{eq:FL:epsilon_doublesum}). 
This would prevent the divergence of Eq.~(\ref{eq:FL:Eq_derivative}).

In contrast, the energies are well behaved and we have $\varepsilon_\mu(q) \rightarrow 0$ for $q \rightarrow 0$ (see Eq.~(\ref{eq:FL:epsilon_doublesum})). 
To compare to our \abinitio results in Sec.~\ref{sec:chResults} (Fig.~\ref{fig:TM_Dvec_AtRes}), we evaluate for the rest of this appendix the energy at a fixed wave vector $q_0 = 0.05 \, \frac{2\pi}{a_y}$ (corresponding to a pitch of \einheit{5.6}{nm}) as magnetic state and calculate $E_\mrm{DM}^\mu$ numerically. 
For this $q_0$, the evaluation of the sum in Eq.~(\ref{eq:FL:epsilon_doublesum}) in a supercell containing 4000 unit cells in the $\pm y$-direction yields well converged results. 
We notice, however, that the convergence depends on the value of $q_0$, \ie, that a lower value of $q_0$ would require a larger number of unit cells in order to reach convergence. 
The atom-resolved contributions to the D-vector are then approximated by the finite difference,
\beq
  \vcrm{D}_\mu \approx       \text{const}
                       \cdot \frac{\varepsilon_{\mu}(q_0)}{q_0}
                       \cdot \hat{\vcrm{n}}_{\mu} \punkt
  \label{eq:FL:Eq_derivative_FD}
\eeq
These values compare to the atom-resolved spiralization from our first-principles calculations presented in Fig.~\ref{fig:TM_Dvec_AtRes} and discussed in Sec.~\ref{sec:chResults} of the main text. 
Here, we only discuss the results predicted by the model:

The type of the $3d$ atom (\ie, Mn, Fe, or Co) only enters the prefactor in Eq.~(\ref{eq:FL:Eq_derivative_FD}) through the parameter $\Gamma$. 
Up to a sign, the directions $\hat{\vcrm{n}}_{\mu}$ (visualized by arrows in Fig.~\ref{fig:TM_Dvec_AtRes}(d) in the main text) are independent of the type of magnetic chain $3d$ atoms. 
The dependence of the DMI strength on the substrate atom is governed by $\varepsilon_\mu$ for a fixed $q_0$. 
For the results in this paper that utilize the Fert-Levy model (see Eq.~(\ref{eq:FL:epsilon_doublesum})) we use the following parameters: 
$2\pi k_\mrm{F}^{-1}=2~\mrm{nm}$ as given by Ref.~\onlinecite{ZhouNatPhys10} and $\delta_2=\frac{\pi}{10} Z_d$, where $Z_d=9.4$ gives the number of $d$-electrons.\cite{FL80}

\begin{figure}[tb]
 \centering
   \includegraphics[width=0.45\textwidth]{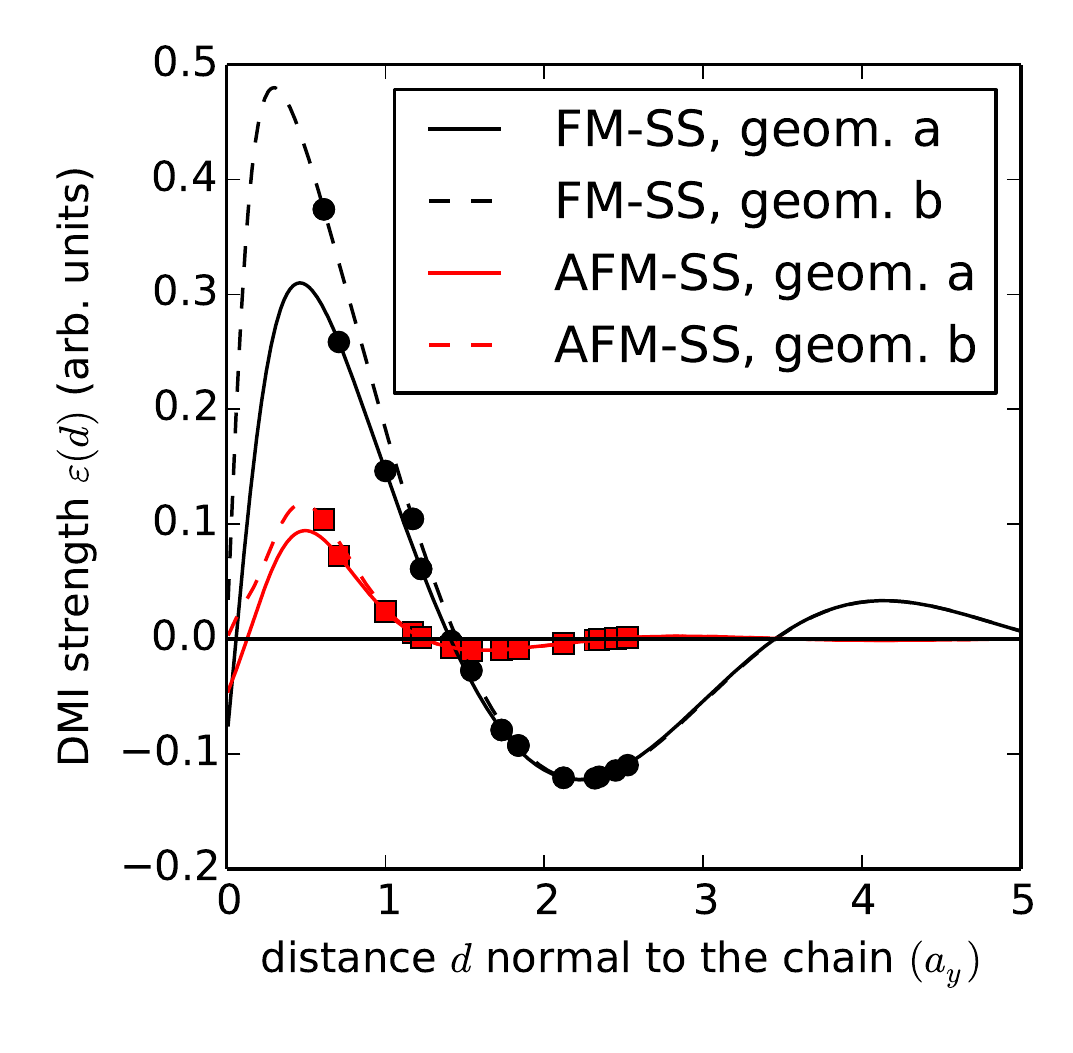}
 \caption{(color online) 
          Dependence of the DMI energy originating form a Pt atom, as function of the distance $d$ of the Pt atom normal to the chain. The nearest-neighbor distance is denoted by $a_y$.
          For the magnetic chain, we distinguish spin spirals of ferromagnetic (FM-SS) or antiferromagnetic (AFM-SS) short-range order, evaluated at a fixed $q_0 = 0.05 \, \frac{2\pi}{a_y}$.
          Points and squares highlight the actual positions of atoms in the Pt(664) unit cell. 
          Two different geometries need to be considered. 
          For a description of the geometries see text.}
 \label{fig:FL_radialdependence}
\end{figure}

In Fig.~\ref{fig:FL_radialdependence}, we analyze $\varepsilon_\mu$ as function of the distance $d$ of Pt atom $\mu$ to the chain for two different cases: that the spin-spiral is of 
($i$)  ferromagnetic short-range order (FM-SS) as the case for Fe and Co, or of 
($ii$) anti-ferromagnetic type (AFM-SS) as in the case of Mn (see circles and squares, respectively). 
More precisely, $\varepsilon(d_\mu)$ is a function of the distance $d$ of atom $\mu$ to the chain if we distinguish two geometrical cases: 
(a) that the projection of the position of atom $\mu$ coincides with the position of a $3d$ atom, or 
(b) that this projection is in the middle of two $3d$ atoms (see Fig.~\ref{fig:FL_radialdependence}). 
This distance dependence is indicated by solid and broken lines, respectively. 
We observe typical RKKY-like oscillations that decay approximately as $1/d^2$. 
In total, we find that for the AFM case the DMI strength is smaller and decays faster with distance than for the FM case. 
As a result, in AFM-SS the first maximum determines the overall DMI strength nearly alone.
Moreover, the period length of the oscillations is nearly by a factor 2 larger in FM-SS ($\einheit{4.5}{a_y}$ and $\einheit{2.5}{a_y}$ for FM-SS and AFM-SS, respectively).

The disagreement between the Fert-Levy model and the \abinitio results for Co (see Figs.~\ref{fig:TM_Dvec_AtRes}(c) and \ref{fig:TM_Dvec_AtRes}(d)) cannot be resolved by adjusting the parameters for Co or the different Pt atoms (\eg, the phase shifts or Fermi wavevector) since for the most important Pt atoms, those next to Co, the direction of $\Dvec$ does not coincide at all with the model of Fert and Levy, where the direction is exclusively determined by geometry. 
Maybe in the case of Co the interaction between the upper and lower Co chains, which is included in our \abinitio calculations, but neglected in the Fert-Levy model contributes to this difference. 
This and the extension of the model to films and heterostructures will be a matter of future investigations.

\bibliography{bibliography}

\end{document}